\documentclass[aps,pre,reprint,twocolumn,superscriptaddress,floatfix,nofootinbib]{revtex4-1}
\usepackage{amsmath,amssymb,color,comment,physics}
\usepackage[makeroom]{cancel}
\usepackage[caption=false]{subfig}
\usepackage{mathrsfs}
\usepackage{graphicx}
\usepackage{subfig}
\usepackage[countmax]{subfloat}
\usepackage[english]{babel}
\usepackage[bookmarks=true,colorlinks,linkcolor=OrangeRed,urlcolor=NavyBlue,citecolor=RoyalBlue]{hyperref}
\usepackage[dvipsnames]{xcolor}
\usepackage{dsfont}

\newcommand{\cref}[1]{Eq.~\eqref{#1}}
\newcommand{\Fig}[1]{Fig.~\ref{#1}}
\newcommand{\Sec}[1]{Sec.~\ref{#1}}

\newcommand{\bk}{\mathbf{k}}
\newcommand{\bx}{\mathbf{x}}
\newcommand{\bxi}{\boldsymbol{\xi}}
\newcommand{\br}{\mathbf{x}}
\newcommand{\bphi}{\boldsymbol{\Phi}}

\newcommand{\st}{{\rm st}}
\newcommand{\eq}{{\rm eq}}
\newcommand{\eff}{{\rm eff}}

\def\diff{\mathrm d}

\definecolor{mypink1}{rgb}{0.858, 0.188, 0.478}
\definecolor{mygold}{rgb}{0.93,0.49,0.13}

\makeatletter
\@ifundefined{textcolor}{}
{%
 \definecolor{BLACK}{gray}{0}
 \definecolor{WHITE}{gray}{1}
 \definecolor{RED}{rgb}{1,0,0}
 \definecolor{GREEN}{rgb}{0,1,0}
 \definecolor{BLUE}{rgb}{0,0,1}
 \definecolor{CYAN}{cmyk}{1,0,0,0}
 \definecolor{MAGENTA}{cmyk}{0,1,0,0}
 \definecolor{YELLOW}{cmyk}{0,0,1,0}
}
\begin{document}
\title{Quantum aging and dynamical universality in the long-range $O(N\to\infty)$ model}
\author{Jad C.~Halimeh}
\affiliation{INO-CNR BEC Center and Department of Physics, University of Trento, Via Sommarive 14, I-38123 Trento, Italy}
\affiliation{Kirchhoff Institute for Physics, Ruprecht-Karls-Universit\"{a}t Heidelberg, Im Neuenheimer Feld 227, 69120 Heidelberg, Germany}
\affiliation{Institute for Theoretical Physics, Ruprecht-Karls-Universit\"{a}t Heidelberg, Philosophenweg 16, 69120 Heidelberg, Germany}

\author{Mohammad F.~Maghrebi}
\affiliation{Department of Physics and Astronomy, Michigan State University, East Lansing, Michigan 48824, USA}

\begin{abstract}
Quantum quenches to or near criticality give rise to the phenomenon of \textit{aging}, manifested by glassy-like dynamics at short times and far from equilibrium. The recent surge of interest in the dynamics of quantum many-body systems has rejuvenated interest in this phenomenon. Motivated by the ubiquitous long-range interactions in emerging experimental platforms, it is vital to study quantum aging in such settings. In this work, we investigate the dynamical universality and aging in the $d$-dimensional $O(N)$ model with the long-range coupling $1/x^{d+\sigma}$ and in the mean-field limit $N\to\infty$ that allows an exact treatment. An immediate consequence of long-range coupling is the emergence of nonlinear light cones. We focus on the correlation and response functions, and identify a rich scaling behavior depending on how the corresponding space-time positions are located relative to each other, via a \textit{local light cone}, and to the time of the quench via a global \textit{quench light cone}. We determine the initial-slip exponent that governs the short-time dependence of two-point functions. We highlight the new qualitative features of aging due to the long-range coupling, in particular in the region outside the light cones. As an important consequence of long-range coupling, the correlation function decays as $1/x^{d+\sigma}$ outside the quench light cone while increasing polynomially with the total time after quench. This is while, for short time differences, the two-time response function ``equilibrates'' at \textit{all} distances even outside this light cone. Our analytic findings are in excellent agreement with exact numerics, and provide a useful benchmark for modern experimental platforms with long-range interactions.
\end{abstract}

%\date{\today}
\maketitle

\tableofcontents

\section{Introduction}
Universality and scaling have long fascinated physicists as they allow a classification of physical systems based on the universal behavior of their macroscopic properties in the vicinity of a critical point, independently of their microscopic details. In equilibrium physics, universality is well established due to a large body of work from Landau's theory of continuous phase transitions \cite{Cardy_book,Ma_book,Sachdev_book} to Wilson's celebrated renormalization group theory \cite{Wilson1971a,Wilson1971b,Wilson1972,Wilson1974,Wilson1975}. In the vicinity of a continuous phase transition, critical behavior arises where only dimensionality, range of interactions,
and the underlying symmetry become relevant.

Out of equilibrium, the theoretical framework of universality and scaling has been extended to classical systems, and distinct dynamical universality classes have been identified from a dynamical renormalization group theory \cite{hohenberg1977theory,Taeuber_book}, but its further extension into the quantum realm is far from complete. What makes this topic particularly rich is that dynamical universality not only depends on dimensionality, range of interactions, and global symmetry, but also on conservation laws and integrability or lack thereof \cite{Moeckel2008,Mazets2008,Rigol2009a,Rigol2009b}. At the same time, quantum many-body systems far from equilibrium have been at the focus of intense research in recent years, where various notions of dynamical criticality have received much attention \cite{Zvyagin2016,Mori2017,Heyl_review}. With long-range interactions, quench dynamics exhibits rich physics such as long-lived prethermal states due to constrained dynamics \cite{Vanderstraeten2018,Halimeh2017a,Halimeh2017b},  super-ballistic propagation of information and faster-than-linear light cones \cite{Hauke2013,Foss-Feig15,Maghrebi2016,Vanderstraeten2018}, and nonequilibrium critical behavior \cite{Titum20}. On the other hand, experimental advances in modern quantum simulators have allowed unprecedented control of real-time dynamics, and have led to the observation of various hitherto theorized phenomena such as gauge-theory dynamics \cite{martinez2016,Goerg2019,Schweizer2019,Mil2019,Yang2020}, many-body localization \cite{schreiber2015,smith2016,Choi2016}, time crystals \cite{choi2017observation,zhang2017observation,Rovny2018,Smits2018}, dynamical phase transitions \cite{jurcevic2016,Flaeschner2017,Zhang2017,Smale2019,Yang2019,Tian2020}, many-body dephasing \cite{kaplan2020manybody}, and prethermalization \cite{Gring2012,Langen2015,Neyenhuis2017,2019Singh}.

An intriguing aspect of critical dynamics is the short-time universality that
occurs after a quench to or near to a critical point \cite{Janssen1989,calabrese_ageing_2006, Henkel11}. While criticality is usually reserved for long distances and times, the short-time universality, also called ``aging'' \cite{Struik1978}, becomes manifest well before the system approaches a steady (possibly thermal) state at late times; this is particularly appealing as available experiments are limited by evolution times.
This notion has originated from the theory of boundary critical phenomena that is concerned with critical properties of magnets, binary alloys, and fluids near surfaces \cite{Diehl81,Diehl86}; aging is just its dynamical counterpart that occurs in space-time with the time slice at $t=0$ playing the role of a surface. In contrast to near-equilibrium dynamics \cite{hohenberg1977theory}, aging leads to truly out-of-equilibrium critical behavior and even a new critical exponent known as the initial-slip exponent $\theta$ with no counterpart in or near equilibrium. This exponent governs the short-time dynamics of the order parameter as well as the critical behavior of the two-time ($t$ and $t'<t$) correlation and response functions. Since the system has not yet equilibrated in the prethermal regime and time-translation invariance is broken, the two-time functions depend not only on the time difference ($t-t'$) but also on the \textit{age} or \textit{waiting time} $t'$:
the older a system is the slower it responds \cite{calabrese_ageing_2006}---a phenomenon that is reminiscent of aging in spin and structural glasses \cite{BOUCHAUD,Vincent1997,Young_book}.
Aging behavior has also been observed experimentally \cite{Nagaya2002,Nagaya2004}, e.g., in magnetic systems quenched from their high-temperature phase to or below their critical point \cite{Bray1994,domb1972phase,calabrese_ageing_2006}.
Studies of aging have only begun to reach the quantum domain where a quench to the critical temperature is replaced by a quantum quench to a (dynamical) critical point \cite{Gambassi2011,schmalian_2014,buchhold_lutt_2015,ciocchetta_2015,Chiocchetta2017}.

Aging in long-range interacting systems has also been studied in various contexts such as the Ginzburg-Landau model \cite{Chen2000}, spherical model \cite{Baumann2007}, anisotropic cubic systems \cite{Chen2001}, random Ising chains \cite{Chen2002}, two-dimensional Ising model \cite{christiansen2019non}, and open $d$-dimensional quantum Ising lattices \cite{Halimeh2018}. On the one hand, solving the quench dynamics of generic long-range quantum many-body systems is particularly challenging, because the Hilbert space grows exponentially in system size.
As a consequence, studying aging in such systems requires approximations whose benchmarking is not straightforward. On the other hand, the interplay of dimensionality and range of interactions can lead to diverse dynamical critical behavior that is not directly accessible in long-range one-dimensional many-body chains \cite{weidinger2017dynamical,Hashizume2018,Hashizume2020}. This provides motivation for studying critical dynamics in a solvable model where the range of interactions and dimensionality can be tuned. In this work, we consider the long-range $d$-dimensional $O(N)$ model with power-law coupling $1/x^{d+\sigma}$ in the limit $N\to\infty$ that allows an exact treatment. The equilibrium phase diagram of this model has been recently mapped out \cite{Defenu2017}. We  study the nonequilibrium critical dynamics and aging in the wake of a quantum quench to or below the dynamical critical point. Specifically, we identify the initial-slip exponent $\theta$ as a function of dimensionality and $\sigma$. Dynamical criticality in the wake of a quench has been investigated in the short-range classical \cite{Hase2006,Ebbinghaus2008,Chiocchetta2016a} and quantum \cite{Smacchia2015,Maraga2015,Chiocchetta2016b,Chiocchetta2017} $O(N)$ models, as well as classical variants of this model with long-range coupling \cite{Chen2000,Chen2001,Chen2002,Baumann2007}; see also \cite{Henkel11}. 
Many parallels with the dynamical criticality in these models notwithstanding, the long-range quantum model finds distinctive features, including a nonlinear light cone and nontrivial power-law dependence of correlations on both distance and time even beyond the light cone, as summarized in the next section. 

The rest of the paper is organized as follows: In Sec.~\ref{sec:summary}, we provide a summary of our main findings and particularly the scaling of correlation and response functions in various regimes. In Sec.~\ref{sec:model}, we present the model along with the quench protocols employed in our study. We study the universal dynamics ensuing a critical quench in Sec.~\ref{Sec. Scaling hypothesis}, and introduce two-point functions and provide analytical expressions in momentum space in Sec.~\ref{sec:TwoPointFns}. We present a detailed analysis of the correlation and response functions in position space in Secs.~\ref{sec:critical} and~\ref{sec:subcritical} for quenches to and below the dynamical critical point, respectively, and identify the corresponding initial-slip exponents while comparing to exact numerics. 
We derive the equilibrium and dynamical critical points for the long-range quantum $O(N\to\infty)$ model in Appendix~\ref{sec:CriticalPoint}. For completeness, a detailed study of the classical variant of the long-range model considered here is provided in Appendix~\ref{app:classical}.

\section{Summary of main results}\label{sec:summary}
\begin{figure}[p]
	\centering
	\includegraphics[width=.47\textwidth]{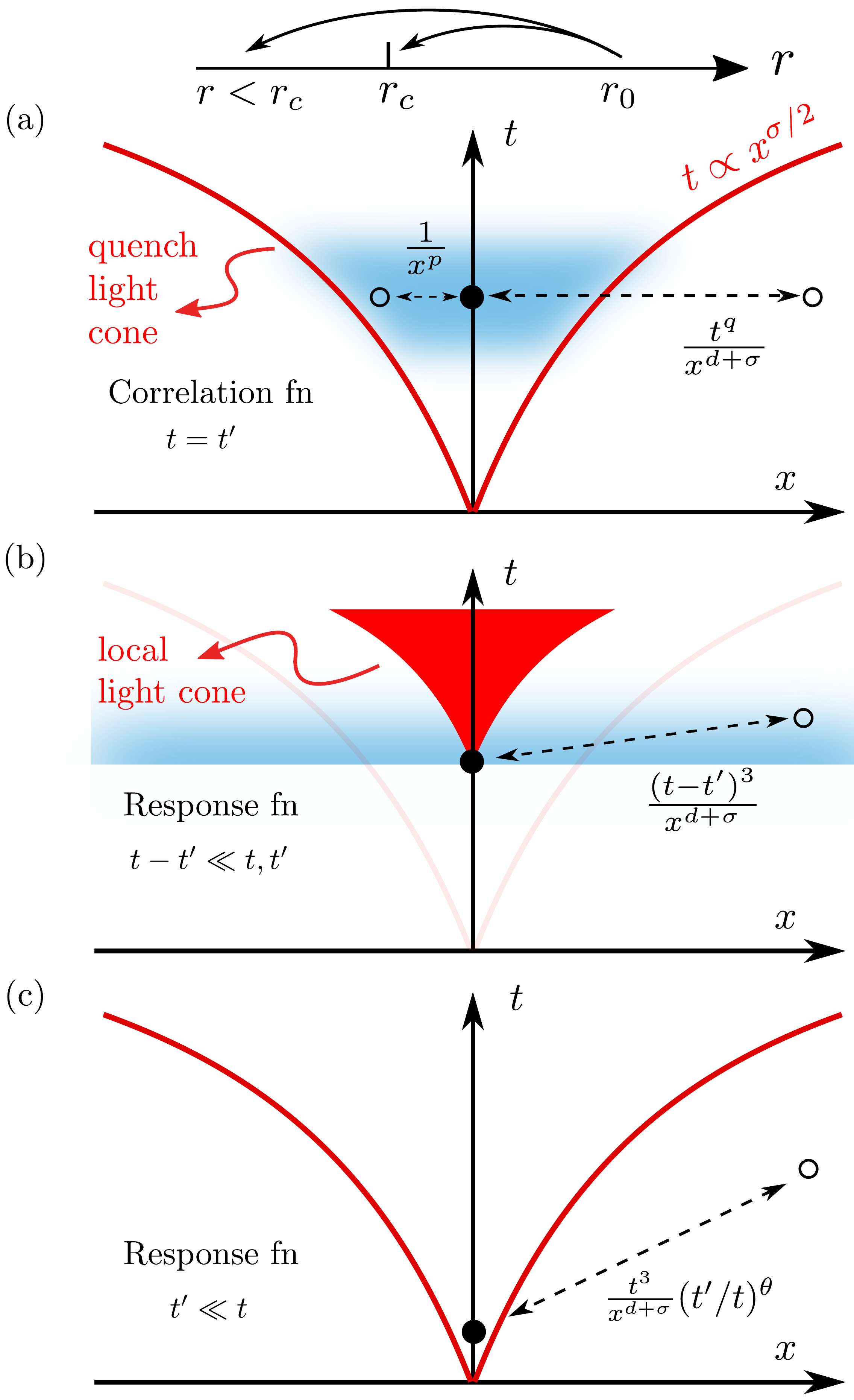}
	\caption{(Color online). Scaling of correlation and response functions after a quantum quench at $t=0$ from a disordered initial state to or below the dynamical critical point ($r\le r_c$) of the $O(N\to \infty)$ model with the long-range coupling $1/x^{d+\sigma}$. 
	Even a quench \textit{below} the dynamical critical point ($r<r_c$) leads to power-law correlations indicative of criticality, but distinct from those in a quench to the dynamical critical point ($r=r_c$). The two-point functions are between space-time points $(x,t)$ represented by an open circle and  $(0,t')$ represented by the solid circle. 
	(a) Equal-time correlation function inside and outside the nonlinear \textit{quench light cone}, $t\propto x^{\sigma/2}$. Well within this light cone, the correlation function shows critical scaling [see \cref{eq:p} for the exponent $p$] and becomes stationary in the highlighted (blue) region. Outside this light cone, it falls off with distance as $1/x^{d+\sigma}$ while increasing with time as a power law [see \cref{eq:q} for the exponent $q$]. (b) The causal response function
	when $t-t'\ll t',t$. In the highlighted (blue) region, the response function has \textit{equilibrated} to its form in equilibrium and depends only on $t-t'$. 
	The response function is best characterized by identifying a nonlinear \textit{local light cone} with $t-t'\propto x^{\sigma/2}$ (filled cone).
	Outside this light cone, the response function decays as $1/x^{d+\sigma}$. The quench light cone is \textit{invisible} in this regime. (c) Aging behavior from the response function when $t' \ll t$. The response function is proportional to $(t'/t)^\theta$, and thus depends on the \textit{age} of the quenched state, with $\theta$ the initial-slip exponent [see \cref{eq:theta}].}
	\label{fig:summary}
\end{figure}

In this work, we consider the quantum $O(N\to\infty)$ model with long-range coupling $\sim1/x^{d+\sigma}$ where $x$ is distance and $d$ is the spatial dimension; cf.~Eq.~\eqref{eq:Ham}. We focus on the critical properties of the transient dynamics as well as the stationary state reached at late times after a quench of the mass $r$ to or below the (dynamical) critical point $r_c$ where critical scaling  behavior emerges. We identify both the correlation and response functions, with the latter characterizing the causal behavior, after the quench; cf.~Eqs.~\eqref{eq:causal}. We find it useful to draw an analogy with light-cone dynamics. Although long-range interactions change the linear nature of the light cone, it will be useful to consider a nonlinear 
light cone, $t\propto x^{\sigma/2}$, which we dub the \textit{quench light cone}. 
Indeed, the dynamical critical exponent is given by $z=\sigma/2$; for short-range coupling (upon inserting $\sigma=2$), we recover $z=1$ indicating the linear-light-cone dynamics.
Let us now consider two space-time points one at $(x,t)$ and a reference point at $(x=0,t')$; see \Fig{fig:summary}. Depending on whether or not the two points are in the same nonlinear light cone starting at $(0,0)$, they exhibit distinct correlations and causal behavior.
In \Fig{fig:summary}(a), we show the correlation function at equal times when $t=t'$. The equal-time correlation function within the quench light cone is given by a power law $1/x^{p}$ with
\begin{equation}\label{eq:p}
    p=
    \begin{cases}
    d/2, & r=r_c, \quad \sigma<d<2\sigma, \\
    \sigma/2, & r<r_c, \quad \sigma<d. \\
    \end{cases}
\end{equation}
The bounds on the dimension specify the lower and upper critical dimensions where the critical behavior is nontrivial (i.e., non-Gaussian); for a quench to $r=r_c$ in any dimension $d>2\sigma$, a Gaussian fixed point governs the critical behavior. In a quench below the dynamical critical point, $r<r_c$, one might expect coarsening dynamics and ordering at long times with the (disconnected) correlation function approaching a constant at long distances; however, this regime too exhibits critical behavior, but one that is distinct from that at $r=r_c$. This behavior has also been viewed as a kind of anomalous coarsening \cite{Chandran13,Sciolla2013,Maraga2015}.
Very interestingly, there is no upper critical dimension in this case and the power-law decay of the correlation function is independent of the dimension. Finally, the highlighted (blue) region schematically denotes the space-time region  $|t-t'|, \,x^{\sigma/2}\ll t,t'$ centered around the reference point $(x=0, t')$ where the correlation function has saturated to a stationary value.
For unequal times ($t\ne t'$), this means that the correlation function (as well as the response function; see below) only depends on the time difference $t-t'$ and not the absolute times. For completeness, we remark that in equilibrium (in the absence of quench), the exponent $p=d-\sigma/2$ at the zero-temperature critical point and $p=d-\sigma$ at a finite-temperature critical point (within the corresponding lower and upper critical dimensions), thus setting the exponent $p$ in the stationary state in \cref{eq:p} apart from those in equilibrium.

Outside the quench light cone ($t\ll x^{\sigma/2}$), the correlation function decays with the same power law as the long-range coupling ($1/x^{d+\sigma}$); however, it also increases in time as $t^q$ with the exponent
\begin{equation}\label{eq:q}
    q=
    \begin{cases}
    2+d/\sigma, & r=r_c, \quad \sigma<d<2\sigma, \\
    1+2d/\sigma, & r<r_c, \quad \sigma<d. \\
    \end{cases}
\end{equation}
Therefore, the correlation function has not saturated to a stationary value in this region. The fact that the correlation function has reached a stationary value inside the quench light cone is consistent with the picture given in terms of oppositely moving quasiparticles in one-dimensional lattice models as well as quantum and conformal field theories \cite{Calabrese_05-2,Calabrese05}. 
The important difference here is that, due to long-range coupling, the light cone is nonlinear and furthermore the correlations outside of it do not decay exponentially but rather vary algebraically with space and time.

In considering the response function after a global quench, one is immediately led to consider another (nonlinear) light cone that defines the causal relation between the two space-time points considered, which we dub the \textit{local light cone}; see the (filled) nonlinear light cone starting from $(x=0,t')$ in \Fig{fig:summary}(b). The causal response function strongly depends on whether or not one point is within the local light cone emanating from the other. We shall focus on the behavior of the response function in the region outside the local light cone.
For short-range models, the response function decays exponentially outside the linear light cone \cite{Lieb72} and is identically zero for relativistic field theories due to causality. On the other hand, in the presence of long-range coupling, the response function (at fixed time) often decays with the same power law as the long-range coupling, in harmony with the Hastings-Koma bound \cite{Hastings05}. This should be expected since any perturbation immediately affects a distant point with a strength that decays with the long-range coupling. Indeed, we find the same behavior within the nonequilibrium dynamics of our model; see \Fig{fig:summary}(b,c). When $t-t' \ll t,t'$, the response function becomes identical to the critical response function in equilibrium that itself is independent of temperature, a fact that is true inside or outside the local light cone and for a quench to or below the dynamical critical point; see the highlighted (blue) region in \Fig{fig:summary}(b). In other words, the quench light cone is \textit{invisible} to the response function in this regime. Also note that, by definition, the causal response function vanishes for $t'> t$. 

Finally, we consider the response function in the most interesting regime where $t'\ll t$; see \Fig{fig:summary}(c). This limit exposes the critical dynamics at short times and contains new information, and a new critical exponent, beyond that at long times. In fact, the response function, beside an overall dependence on $x$ and $t$, becomes proportional to $(t'/t)^\theta$ with $\theta$ the so-called initial-slip exponent \cite{Janssen1989,calabrese_ageing_2006,Henkel11}; the overall dependence on the age of the system is called aging. In the context of the quantum model considered here, we find
\begin{equation}\label{eq:theta}
    \theta=
    \begin{cases}
    1-d/(2\sigma), & r=r_c, \quad \sigma<d<2\sigma, \\
    3/2-d/\sigma, & r<r_c, \quad \sigma<d. \\
    \end{cases}
\end{equation}
For a quench to $r=r_c$, we have $\theta>0$, while for a quench below the dynamical critical point, $r<r_c$, this exponent can become negative at sufficiently high dimensions, namely when $d>3\sigma/2$.
In a similar fashion, aging and the initial-slip exponent become manifest in the unequal-time correlation function in the same limit, $t'\ll t$.

\section{Model and quench}\label{sec:model}
We consider the $d$-dimensional long-range $O(N)$ model given by the Hamiltonian

\begin{align}\label{eq:Ham}
H(r,u)=\int \diff^d\bx \left[\frac{1}{2}\boldsymbol{\Pi}^2-\frac{1}{2}\bphi\cdot\nabla^\sigma\bphi+\frac{r}{2}\bphi^2+\frac{u}{4!N}(\bphi^2)^2\right],
\end{align}
with the $N$-component bosonic field $\bphi=[\Phi_1(\bx),\ldots,\Phi_N(\bx)]$ and the canonically conjugate momentum $\boldsymbol{\Pi}=[\Pi_1(\bx),\ldots,\Pi_N(\bx)]$, which satisfy the commutation relation $[\Phi_m(\bx),\Pi_n(\bx')]=i\delta(\bx-\bx')\delta_{mn}$. Here we have used the shorthand $\nabla^\sigma$ to denote the long-range coupling $\sim 1/x^{d+\sigma}$ for $0<\sigma<2$; we use $x=|\bx|$ for convenience (similarly for $\bk$). More precisely, the long-range coupling in momentum space is given by $\frac{1}{2}\int \frac{\diff^d\bk}{(2\pi)^d} k^\sigma |\bphi(\bk)|^2$ where $\bphi(\bk)$ is the Fourier transform of the field.
By setting $\sigma=2$ in Eq.~\eqref{eq:Ham}, one retrieves the short-range $O(N)$ model. The squared operators denote the inner product of the vector fields. Finally, $u>0$ denotes the strength of the nonlinear interactions, while $r$ defines the (bare) \textit{mass} of the bosonic field. For $\sigma>2$, the momentum dependence $\sim k^2$ due to the short-range coupling is dominant and the long-range coupling may be neglected insofar as the critical properties are concerned. In this work, we restrict ourselves to $\sigma<2$.

We shall prepare the system in the ground state of the initial Hamiltonian $H(\Omega_0^2,0)\equiv H_0$ with the initial mass $r_0=\Omega_0^2$ and then quench the system to the final Hamiltonian $H(r,u)$ at $t=0$.\footnote{One can start with a nonzero initial $u$, but this will only lead to an inconsequential renormalization of $\Omega_0^2$ and does not change the dynamical criticality.} This quench protocol has been extensively studied in the short-range limit \cite{Chandran13, Sciolla2013,Smacchia2015, Maraga2015, ciocchetta_2015, Chiocchetta2016b,Chiocchetta2017}. 
Starting from the disordered phase, the initial state inherits the $O(N)$ symmetry. Together with the same symmetry of the post-quench Hamiltonian, the state at all times respects the symmetry. Specifically, this implies that the correlation function $\langle \Phi_m\Phi_n\rangle$ vanishes for cross correlations ($m\ne n$) and is independent of the field component $n$ for diagonal correlations ($m=n$). In the limit $N\to \infty$, we exploit the mean-field character of the model to replace the quartic interactions by $\lim_{N\to \infty}\frac{1}{N}(\bphi^2)^2= 2\langle\phi^2\rangle \bphi^2$ where $\phi$ represents any component of the field $\bphi$ due to the symmetry. The dynamics is then governed by a quadratic Hamiltonian for each field $\phi$ (dropping the component index) and its conjugate momentum $\Pi$ as
\begin{subequations}
\begin{equation}
    H_{\rm eff}(t)=\frac{1}{2}\int \diff^d\bx \left[\Pi^2-\phi\nabla^\sigma\phi+r_{\rm eff}(t)\phi^2\right],
\end{equation}
where
\begin{equation}\label{Eq. r_efff}
    r_{\rm eff}(t)=r+\frac{u}{6}\langle \phi^2(\bx,t)\rangle.
\end{equation}
\end{subequations}
These equations define a self-consistent set of equations where the field $\phi$ evolves under the Hamiltonian $H_{\rm eff}$ whose parameter $r_{\rm eff}$ itself depends on the field.
We shall expand the fields in terms of creation and annihilation operators $a_\bk$ and $a_\bk^\dagger$ as
\begin{equation}\label{Eq. f_bk}
    \phi_\bk(t)=f_\bk(t)a_\bk +f^*_\bk(t)a^\dagger_{-\bk},
\end{equation}
which diagonalize the initial Hamiltonian, $H_0=\int_\bk \omega_{0\bk}(a^\dagger_\bk a_\bk+1/2)$ with $\omega_{0\bk}=\sqrt{k^\sigma+r_0}$; for notational convenience, we have defined $\int_\bk\equiv \int \diff^d \bk/(2\pi)^d$. The Heisenberg equation of motion will then determine the evolution of the time-dependent coefficients $f_\bk(t)$ via
\begin{subequations}\label{Eq. r_eff}
\begin{align}
    \ddot f_\bk+[k^\sigma+r_{\rm eff}(t)] f_\bk=0, \label{Eq. r_eff-1} \\
    r_{\rm eff}(t)=r+\frac{u}{6}\int_\bk |f_\bk(t)|^2, \label{Eq. r_eff-2}
\end{align}
\end{subequations}
where the integral in the last equation is up to a high-momentum cutoff $\Lambda$, i.e., $|\bk|\le \Lambda$. Such a cutoff makes the theory well-defined at short distances and does not affect it at all for $k\ll\Lambda$, and therefore it will not affect the critical properties of the system at long distances. Hard cutoffs are however pathological when it comes to dynamics, as they lead to spurious oscillations in observables that hide their universal long-time behavior \cite{Maraga2015}. Accordingly, they should be replaced with a soft cutoff that can be introduced by a function $h(k/\Lambda)$ which smoothly, but quickly decays when its argument is large compared to one. In all numerical simulations in this work, we have used $h(k/\Lambda)=\exp\{-k^2/(2\Lambda^2)\}$; we have checked that 
other choices of a soft cutoff such as $h(k/\Lambda)=\exp\{-k/\Lambda\}$ do not alter our main results. Finally, the equation of motion should be supplemented with the initial conditions, $f_\bk(0)=1/\sqrt{2\omega_{0\bk}}$ and $\dot f_\bk(0)=-i\sqrt{\omega_{0\bk}/2}$. We further assume that the system is initially deep in the disordered state, $\omega_{0\bk}\approx \Omega_0$. In this case, the initial conditions simply become $f_0=1/\sqrt{2\Omega_0}$ and $\dot f_0=-i\sqrt{\Omega_0/2}$, independent of the momentum $\bk$. 

The numerical simulation of the equations of motion is achieved by using the Delambre-St\"ormer-Verlet method \cite{Verlet1967}. In all simulations, we choose $u=80$, $\Omega_0=10$, and $\Lambda=\pi/2$; we have checked that the universal results reported here hold for other choices of these parameters.
For the most demanding simulations, we achieve convergence with a time step of $\delta t=10^{-4}$ and a discretization of $\delta k= 10^{-5}\pi$ within the range $k \in [0, 4\pi]$.

\section{Critical quench: Scaling hypothesis}\label{Sec. Scaling hypothesis}
In this section, we consider a critical quench, i.e., a scenario where the long-time state is critical; this is determined by the condition $r_{\rm eff}(t)\to 0$ as $t\to \infty$. 
Our aim is to investigate the critical state reached at late times, but also identify the universal dynamics at short or intermediate times before approaching the stationary state. The critical dynamics is determined by a scaling ansatz that makes the equation of motion scale invariant. An inspection of Eq.~\eqref{Eq. r_eff-1} reveals that a scaling solution is achieved only if the effective mass decays with time as
\begin{equation}\label{Eq. Scaling ansatz}
    r_{\rm eff}(t)=\frac{a}{t^2},
\end{equation}
where $a$ is a dimensionless constant that should be determined from our self-consistent analysis. Notice that this term scales in the same way as the second-order time derivative, and thus does not introduce an additional time scale. We should stress however that the scaling solution is expected to emerge only after a fast transient evolution over a time scale that is controlled by the short-wavelength cutoff (set by the momentum cutoff).

Inserting the scaling solution \eqref{Eq. Scaling ansatz} in the equation of motion \eqref{Eq. r_eff-1} and defining a scaling function $f_\bk(t)\equiv g_\bk(k^{\sigma/2}t)$, we obtain an equation for $g(x)$ as 
\begin{align}
     g''_\bk+\left(1+\frac{a}{x^2}\right)g_\bk=0.
\end{align}
A general solution to this equation takes the form
\begin{equation}\label{g_bk}
    g_\bk(x)=\sqrt{x}\left[A_\bk J_\alpha(x)+B_\bk J_{-\alpha}(x)\right],
\end{equation}
where $\alpha=\sqrt{1/4-a}$; the coefficients $A_\bk$ and $B_\bk$ in general depend on $\bk$. Apart from the dependence on $\sigma$ in the argument of the function $g$ (i.e., $x\equiv k^{\sigma/2}t$), the above equations are identical to those obtained for the short-range coupling (upon inserting $\sigma=2$) \cite{Maraga2015, Chiocchetta2016b}. Indeed, the subsequent analysis in this section parallels that of short-range interactions, although with modifications following the original work of Janssen \textit{et al.} \cite{Janssen1989}.

To describe the full solution, we must determine the coefficients $A_\bk$ and $B_\bk$ by inspecting the initial conditions. To this end, one can expand Eq.~\eqref{g_bk} at short times. Given the $\bk$-\textit{independence} of the initial values of $f_\bk$ and $\dot f_\bk$, one can show that the coefficients $A_\bk$ and $B_\bk$ should scale with $k$ as
\begin{equation}
\begin{split}
    A_\bk\sim A\, k^{-(1/2+\alpha)\sigma/2}, \\
    \quad B_\bk\sim B\, k^{-(1/2-\alpha)\sigma/2},
\end{split}
\end{equation}
with $A$ and $B$ constants independent of $\bk$. With the milder divergence of the coefficient $B_\bk$ with $k$, one may expect that its contribution would be less relevant in the scaling regime past the short transient time. Indeed, this is confirmed by numerical simulation (see also \cite{Maraga2015}), and, together with Eqs.~(\ref{Eq. r_eff-2},\ref{g_bk}), leads to the self-consistent equation 
\begin{align}\label{eq:r-efff}
\begin{split}
    r_{\rm eff}(t)&=r+ \frac{u}{6}K_d |A|^2  t \int_0^{1} \!\!\diff k\, k^{d-1-\sigma \alpha}J_{\alpha}^2(k^{\sigma/2}t).
\end{split}
\end{align}
Note that we have written the integral $\int {\diff^d} \bk/(2\pi)^d \cdots=K_d \int \diff k \cdots$ where $K_d=\Omega_d/(2\pi)^{d}$ with the spherical angle $\Omega_d=2\pi^{d/2}/ \Gamma(d/2)$. Here and in the subsequent analytical treatment, we set $\Lambda=1$ for ease of notation (though $\Lambda=\pi/2$ in numerical simulations).

Now, for a critical quench, the value of $r$ should be chosen such that $\lim_{t\to \infty}r_{\rm eff}(t)=0$, ensuring a critical state at late times. Using this fact, we can rewrite the above equation together with the scaling ansatz in Eq.~\eqref{Eq. Scaling ansatz}, and  upon a change of variable $z=k^{\sigma/2}t$, as 
\begin{equation}\label{eq:self-cons-a}
    a=\frac{uK_d|A|^2 }{3\pi\sigma}t^{3+2\alpha-\tilde d} \int_0^t \diff z\, z^{\tilde d-2-2\alpha} \left[\pi z J_{\alpha}^2(z) -1\right]. 
\end{equation}
Here, we have defined  $\tilde d=2d/\sigma$. Apart from the coefficient of the integral (only affecting the critical value of $r, u$), the dependence on the exponent $\sigma$ and the dimensionality $d$ is given through the combination $2d/\sigma$ which appears as an exponent in the above integral. This is particularly convenient as it brings the integral into the same form as the short-range case ($\sigma=2$) simply by changing $d \to \tilde d\equiv 2d/\sigma$.
We also remark that the expression in the bracket in the above equation goes to zero (up to a highly oscillatory part) as $z\to \infty$. 

Equation~(\ref{eq:self-cons-a}) defines a self-consistent criterion to determine $a$ (the rhs depends on $a$ through $\alpha$). It is convenient to write the time integral in this equation as $\int^t_0 =\int_0^\infty- \int_t^\infty$. 
The dominant contribution is due to the first integral ($\int_0^\infty$) whose coefficient grows in time\footnote{To be more precise, the dependence of the rhs on time depends on $\alpha$ and $d$; however, one can verify that it indeed grows in time for the values of $\alpha$ obtained later in this section.}  while the lhs is constant. We thus require the condition
\begin{equation}\label{eq:condition}
\int_0^\infty \diff z\, z^{\tilde d-2-2\alpha} \left[\pi z J_\alpha^2(z) -1\right]=0.
\end{equation}
We will evaluate this integral shortly to determine $\alpha$. Before doing so, we note that the remaining integral ($\int_t^\infty$) provides the next order in the expansion of \cref{eq:self-cons-a} from which we can determine the \textit{fixed-point value} of $u$. Let us first expand the expression in the bracket for $z\gg 1$:
\begin{equation}
       \pi z J_\alpha^2(z) -1 \sim \mbox{oscill.~terms}+\frac{4\alpha^2-1}{8z^2} + {\cal O}(1/z^3). \nonumber
\end{equation}
The  oscillatory terms start at the order of ${\cal O}(1/z)$; however, one can verify that they do not contribute to the final answer due to their highly oscillatory nature. The next term in the expansion is proportional to $1/z^2$, from which we can obtain the fixed-point value of the coupling constant as 
\begin{equation}\label{eq:u*}
    u_* =\frac{24\pi \sigma}{K_d|A|^2}\frac{(\tilde d-3-2\alpha)a}{4\alpha^2-1}.
\end{equation}
This equation fixes the value of $u$ given the (yet unknown) value of $a$ and $\alpha$. Most importantly, we shall require $u_*\ge0$, which will put important constraints on the space of possible solutions and will determine the upper critical dimension, as we will see shortly. 

We now return to the condition in \cref{eq:condition} which dictates\footnote{The integral in \cref{eq:condition} is not strictly convergent. One can remedy this by introducing a soft cutoff and inspect the asymptotic behavior at long times. However, a consistent solution is simply obtained by computing the integral with a hard cutoff and then analytically continue it to the regime of interest.}
\begin{equation}
    \frac{\Gamma(\tilde d/2)\Gamma(1/2-\tilde d/2+\alpha)}{\Gamma(1-\tilde d/2+\alpha)\Gamma(1-\tilde d/2+2\alpha)}=0.
\end{equation}
We designate a first set of solutions by $\alpha^{(1)}$:
\begin{equation}\label{Eq. alpha}
    \alpha^{(1)}=\frac{\tilde d-2}{4}\quad \longrightarrow \quad a^{(1)}=\frac{\tilde d}{4}\left(1-\frac{\tilde d}{4}\right).
\end{equation}
Given that $\alpha\ge 0$, we have $\tilde d\ge 2$ which thus sets the lower critical dimension, $d_l=\sigma$. Incidentally, the same lower critical dimension arises at finite temperature in equilibrium. Moreover, the fixed-point value of $u$ can be determined from \cref{eq:u*} as
\begin{equation}
    u_*^{(1)} =\frac{3\pi\sigma}{K_d|A|^2}(4-\tilde d).
\end{equation}
Now, the condition that $u_* \ge 0$ requires $\tilde d \le 4$ and thus identifies the upper critical dimensions $d_u=2\sigma$ which again coincides with that of finite-temperature equilibrium. For any $d>2\sigma$, the fixed point becomes Gaussian, and we simply have $\alpha=1/2$ or $a=0$. 

A second set of solutions to \cref{eq:condition} can be found as 
\begin{equation}\label{Eq. alpha'}
    \alpha^{(2)}=\frac{\tilde d-2}{2} \,\,\, \longrightarrow \,\,\, a^{(2)}=\frac{(3-\tilde d)(\tilde d-1)}{4}.
\end{equation}
Again, the condition $\alpha\ge 0$ implies $\tilde d \ge 2$ identifying the same lower critical dimension, $d_l=\sigma$. Similarly, the fixed-point value of the coupling constant can be obtained from \cref{eq:u*} and is given by
\begin{equation}
    u_*^{(2)} =\frac{6\pi\sigma}{K_d|A|^2}.
\end{equation}
Interestingly, we find that $u_*$ is always positive and thus there is no upper critical dimension.

\begin{figure}[t]
	\centering
	\hspace{-.25 cm}\vspace{0.13cm}
	\includegraphics[width=.49\textwidth]{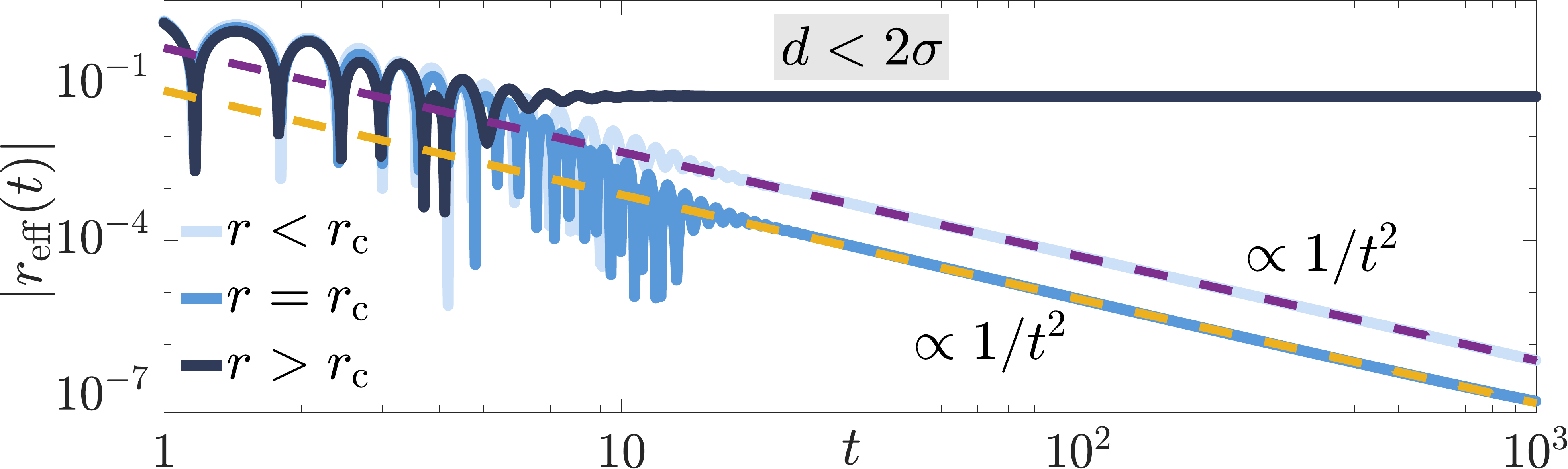}\\
	\hspace{-.25 cm}\vspace{0.13cm}
	\includegraphics[width=.49\textwidth]{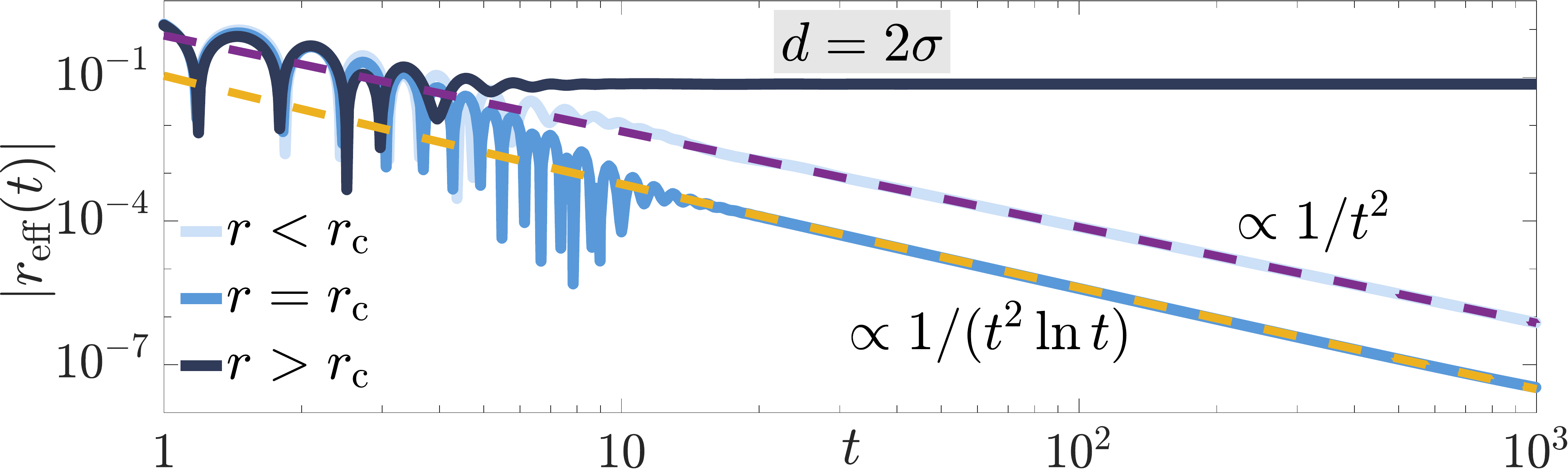}\\
	\hspace{-.25 cm}\vspace{0.13cm}
	\includegraphics[width=.49\textwidth]{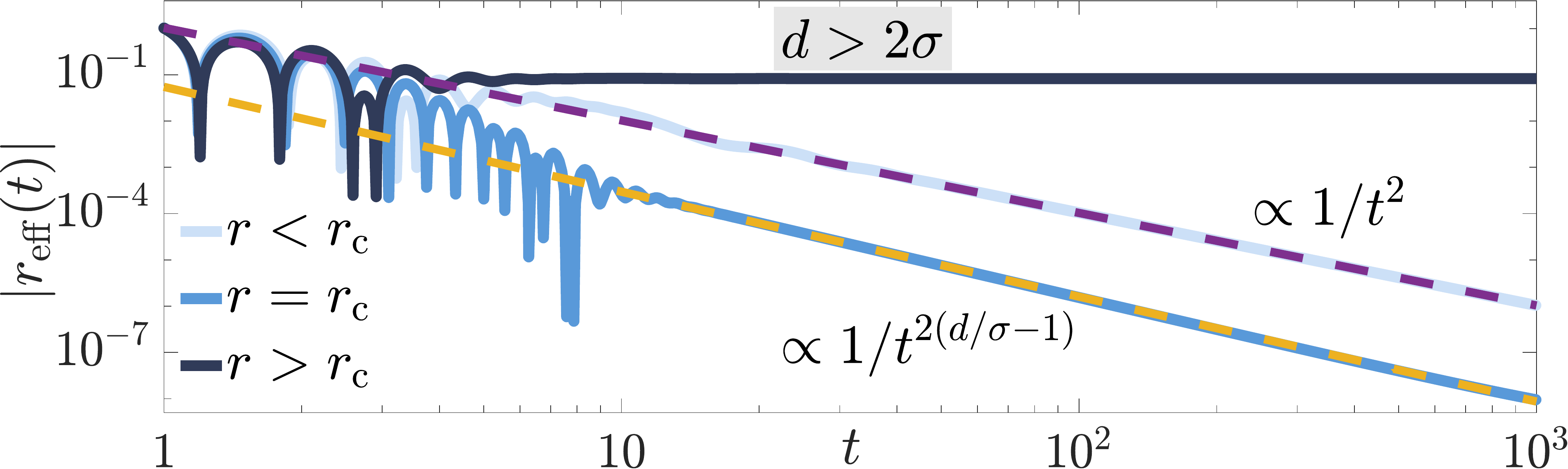}\\
	\caption{(Color online). Time evolution of the effective mass $r_\mathrm{eff}$ for $\sigma=1.5$ in dimensions $d=2.8<d_\mathrm{u}$ (upper panel), $d=3=d_\mathrm{u}$ (middle panel), and $d=3.2>d_\mathrm{u}$ (bottom panel), where each panel shows the time evolution for quenches below, to, and above the dynamical critical point (light to dark blue solid lines, respectively); $d_u=2\sigma=3$ is the upper critical dimension.
	A quench above the dynamical critical point always leads to a nonvanishing effective mass, while a quench to or below the dynamical critical point leads to an effective mass that falls off algebraically with time.}
	\label{fig:mass}
\end{figure}

Given that the first set of solutions is sensitive to the upper critical dimension, it is natural to identify it with a quench to the dynamical critical point $r=r_c$ describing the onset of critical behavior. 
We stress that $r_c$ denotes the \textit{dynamical} critical point characterizing the stationary state long after a quench, and should not be confused with the equilibrium critical point at finite temperature (see \cite{ciocchetta_2015} for a comparison with the thermal critical point).
Any quench above this dynamical critical point ($r>r_c$) leads to a finite effective mass, and hence a finite correlation length. On the other hand, the second set of solutions discussed above is insensitive to the upper critical dimension, therefore we may anticipate that it describes the critical behavior in a quench below the dynamical critical point when $r<r_c$.
Indeed, we can directly verify these points numerically: In \Fig{fig:mass}, we inspect three distinct cases where the dimension is smaller, equal, or larger than $d_u=2\sigma$ and consider three distinct quenches above, to, and below the dynamical critical point $r_c$. 
One can identify the value of $r_c$ numerically, or estimate it via a simple ansatz \cite{Sotiriadis10,Smacchia2015,Maraga2015} described in Appendix \ref{sec:CriticalPoint}. 
In all cases, a quench above the dynamical critical point ($r>r_c$) leads to a finite effective mass $r_{\rm eff}\ne 0$ and the system remains noncritical. A quench to the dynamical critical point ($r=r_c$) for $d<d_u$ leads to the critical scaling $r_{\rm eff} \propto 1/t^2$ while at the upper critical dimension finds a multiplicative logarithmic correction; such logarithms also arise at the upper critical dimension of the classical spherical model, somewhat the classical analog of the model considered here \cite{Henkel11}.
At higher dimensions, $d>d_u$, the scaling hypothesis does not hold, which is expected since the model is governed by a Gaussian fixed point where $a=0$ and $\alpha=1/2$. In this case, the condition in \cref{eq:condition} is not satisfied and the leading term in \cref{eq:r-efff} falls off as $r_{\rm eff}\sim 1/t^{\tilde d-2}$ at long times \cite{Maraga2015}. 
For a quench below the dynamical critical point ($r<r_c$), the scaling hypothesis, $r_{\rm eff} \propto 1/t^2$, is satisfied in all dimensions, indicating the absence of an upper critical dimension, consistent with our analysis. We also note that no critical scaling emerges below or at the lower critical dimension, the latter because the critical value $r_c \to -\infty$ at $d=d_l$; see Appendix~\ref{sec:CriticalPoint}.

Before closing this section, we remark once again that the results obtained in this section closely mirror those of the short-range $O(N)$ model. Indeed, we have seen that upon a change of variable, $d\to \tilde d=2d/\sigma$, Eqs.~(\ref{Eq. alpha},\ref{Eq. alpha'}) become identical to those of the short-range model \cite{Maraga2015,Chiocchetta2016b}. Nevertheless, we shall see in the subsequent sections that the long-range $O(N)$ model has unique features that make it distinct from its short-range counterpart; see also the summary in \Sec{sec:summary}.

\section{Two-point functions}\label{sec:TwoPointFns}
To fully characterize critical properties of our model, we need to identify spatial and temporal correlations. To this end, we introduce the two-point functions
\begin{subequations}\label{eq:causal}
\begin{align}
    iG^R(\bx-\bx',t,t')&=\Theta(t-t') \langle[\phi(\bx,t),\phi(\bx',t')] \rangle, \\
    iG^K(\bx-\bx',t,t')&=\langle\{\phi(\bx,t),\phi(\bx',t')\} \rangle.
\end{align}
\end{subequations}
Here, $G^R$ and $G^K$ represent the response and (symmetrized) correlation functions, respectively; our notation is inspired by the Keldysh field theory conventions where $G^{R/K}$ designate the retarded and Keldysh functions, respectively \cite{kamenev_book}. While the former function describes the causal response to a  perturbation at an earlier time (hence, the Heaviside step function $\Theta$), the latter characterizes correlations; at $t=t'$, the latter function gives the equal-time correlation function. Using spatial translation symmetry, we  have written the two-point functions as a function of $\bx-\bx'$. In momentum space, these functions can be written in terms of $f_\bk$ as [see Eq.~\eqref{Eq. f_bk}]
\begin{subequations}
\begin{align}
    iG^R(\bk,t,t')&=2\Theta(t-t')\Im [f_\bk(t)f_\bk^*(t')], \label{eq:G-RR} \\
    iG^K(\bk,t,t')&=2 \Re[f_\bk(t)f_\bk^*(t')]. \label{eq:G-KK}
\end{align}
\end{subequations}
Our analytical solution for $f_\bk$ together with the corresponding value of $\alpha$ [see Eqs.~\eqref{Eq. alpha} and \eqref{Eq. alpha'}] allow us to obtain analytical solutions for the two-point functions. While there are many parallels with the short-range model \cite{Maraga2015,Chiocchetta2016b}, we will highlight the qualitatively new terms that emerge exclusively due to the long-range coupling.

We shall focus on the equal-time correlation function in which case we simply have $iG^K(\bk,t,t)=2|f_\bk(t)|^2$. This is the same expression that we have encountered before when identifying the self-consistent equation for the effective mass in Eq.~\eqref{Eq. r_eff-2}. 
The above equation together with \cref{g_bk} yield
\begin{align}\label{Eq. C}
 C(\bk,t)&\equiv    iG^K (\bk,t,t) \nonumber \\
 &\approx 2  t |A|^2 k^{-\sigma \alpha} J_\alpha^2(k^{\sigma/2}t).
\end{align}
Here, we have introduced $C$ to denote the equal-time correlation function.
Specifically, the stationary-state correlations are obtained in the limit $t\to \infty$:
\begin{align}\label{eq:C-stationary}
 C_\st(\bk)\equiv \lim_{t\to \infty} C(\bk,t)
 \sim \frac{2|A|^2}{\pi} k^{-\sigma (\alpha+1/2)},
\end{align}
where we have defined the long-time limit of equal-time correlation function $C_\st$ with the subscript st denoting the stationary state. More generally, the correlation function approaches a stationary value once $k^{\sigma/2} t\gtrsim 1$, which describes the region within the quench light cone at \textit{any} finite time. We shall discuss this in further detail in the following sections.

Next we inspect \cref{Eq. C} in the limit of long distances, or $k^{\sigma/2} t\ll 1 $:
\begin{align}\label{eq:C-short-t}
 C(\bk,t)
 \sim \frac{2^{1-2\alpha}|A|^2 }{\Gamma^2(1+\alpha)}  t^{1+2\alpha} \left[1+ \frac{k^\sigma t^2}{2(1+\alpha)}+\cdots\right],
\end{align}
While the first term gives the leading-order contribution in momentum space, it decays quickly in position space. The subleading term however is dominant at long distances and decays algebraically with the same power law as the long-range coupling ($1/x^{d+\sigma}$) outside the quench light cone. Interestingly, the correlation function outside the light cone also increases algebraically in time as we will discuss in detail in the following sections.

One can similarly obtain the response function from \cref{eq:G-RR} together with \cref{g_bk} as 
\begin{align}\label{Eq. G^R}
    &G^R(\bk,t,t')=-\Theta(t-t')\frac{\pi}{2\sin(\pi\alpha)}(tt')^{1/2}\,\times\nonumber \\
    & \left[J_\alpha (k^{\sigma/2}t)J_{-\alpha}(k^{\sigma/2}t')-J_{-\alpha}(k^{\sigma/2}t)J_{\alpha}(k^{\sigma/2}t')\right].
\end{align}
In deriving this equation, we have used the canonical commutation relation $[\phi(\bx,t),\dot\phi(\bx',t)]=i\delta(\bx-\bx')$, which, in momentum space and in terms of the function $f_\bk(t)$, translates into $\Im[f_\bk(t)\dot f_\bk^*(t)]=1$, which in turn yields
\begin{equation}
    \Im(A_\bk B_\bk^*)=-\frac{\pi}{4\sin(\pi\alpha)}\frac{1}{k^{\sigma/2}},
\end{equation}
a fact that we have used to arrive at Eq.~\eqref{Eq. G^R}. 

Next we inspect the response function in several limits. First, we find that at long times, $k^{\sigma/2}t, k^{\sigma/2}t'\gg 1$,
\begin{equation}\label{Eq. G^R infty}
    G^R(\bk,t,t')=-\Theta(t-t')\frac{1}{k^{\sigma/2}}\sin\left[k^{\sigma/2}(t-t')\right].
\end{equation}
This limit is appropriate within the quench light cone, but it is more broadly applicable as we explain shortly.
Notice that the time translation symmetry is restored at long times, hence the dependence on $t-t'$ only. Indeed, in this limit, we recover the response function of a critical quadratic Hamiltonian which is state-independent and is thus independent of temperature. To identify the response function outside the local light cone (but still within the quench light cone), we can expand the above equation for $k^{\sigma/2} (t-t')\ll 1$:
\begin{equation}\label{eq:GR-expanded}
    G^R(\bk,t,t')=\Theta(t-t')[ -(t-t')+\frac{1}{6}k^\sigma (t-t')^3+\cdots].
\end{equation}
This equation follows from \cref{Eq. G^R infty} that itself is derived in the limit $k^{\sigma/2} t, k^{\sigma/2} t'\gg1$; however, it is more generally valid as long as $t-t'\ll t,t',k^{-\sigma/2}$ regardless of $k^{\sigma/2}t$ and $k^{\sigma/2}t'$, as one can directly verify from \cref{Eq. G^R}. 

Next, let us inspect the response function well outside the quench light cone. To this end, we should consider \cref{Eq. G^R} in the limit $k^{\sigma/2}t,\, k^{\sigma/2}t'\ll1$:  
\begin{align}\label{eq:GR-outside-aging1}
    G^R(\bk,t,t')\sim t \,\Phi_\alpha(t'/t) +k^\sigma t^3 \,\Psi_\alpha(t'/t)+\cdots,
\end{align}
where, for convenience, we have defined the functions 
\begin{align}
    \Phi_\alpha(y)&\equiv \frac{1}{2\alpha} \left[y^{1/2+\alpha}-y^{1/2-\alpha}\right], \label{eq:Phi}\\
    \Psi_\alpha(y)&\equiv \frac{y^{\frac{1}{2}-\alpha } }{8 \alpha }\Big\{\frac{y^2-y^{2\alpha}}{1-\alpha}    +\frac{1-y^{2+2\alpha}}{1+\alpha} \Big\}.\label{eq:Psi}
\end{align}
Again, the leading term in \cref{eq:GR-outside-aging1} quickly decays in  position space, while the subleading term provides the leading-order nonanalytical term in $k$ and gives rise to the same power law of the long-range coupling ($\sim 1/x^{d+\sigma}$) at long distances. 
A particularly interesting limit of \cref{eq:GR-outside-aging1} is when $t'\ll t \ll k^{-\sigma/2}$:
\begin{equation}\label{eq:GR-outside-light-cone}
    G^R(\bk,t,t')\sim \frac{t}{2\alpha} \left(\frac{t'}{t}\right)^\theta \left[-1+ \frac{k^\sigma t^2}{4(1+\alpha)}+\cdots\right],
\end{equation}
where we have included the first nonanalytical term in $k$ too. In this equation, we have identified the initial-slip exponent that characterizes the dependence on the ratio $t'/t$:
\begin{equation}\label{eq:th}
    \theta=\frac{1}{2}-\alpha.
\end{equation}

In the following sections, we use the results derived here to investigate the critical behavior in position space and in various limits. In \Sec{sec:critical}, we focus on a quench to the dynamical critical point ($r=r_c$), while, for a clearer presentation, we dedicate \Sec{sec:subcritical} to the study of the critical behavior in a quench below the dynamical critical point ($r<r_c$), although there are many parallels between the two sections.

\section{Quench to the dynamical critical point $r=r_c$}\label{sec:critical}
In this section, we focus on the quench to a dynamical critical point and derive analytical results for the correlation and response functions and identify the initial-slip exponents in various regimes and compare them against exact numerical results. We also derive a $k$-dependent effective temperature that mimics critical correlations at long distances within the stationary state.

\subsection{Correlation function}
We start by considering the initial correlation function at or before $t=0$ due to the long-range coupling \cite{Maghrebi2016}: 
\begin{equation}
    C(\br,t=0)\propto \frac{1}{x^{d+\sigma}}.
\end{equation}
This power law does \textit{not} indicate criticality since $r_0=\Omega_0^2>0$, but is simply inherited from long-range nature of the model \cite{Cardy_book,Maghrebi2016}. 

\begin{figure}[t]
	\centering
	\hspace{-.25 cm}\vspace{0.13cm}
	\includegraphics[width=.49\textwidth]{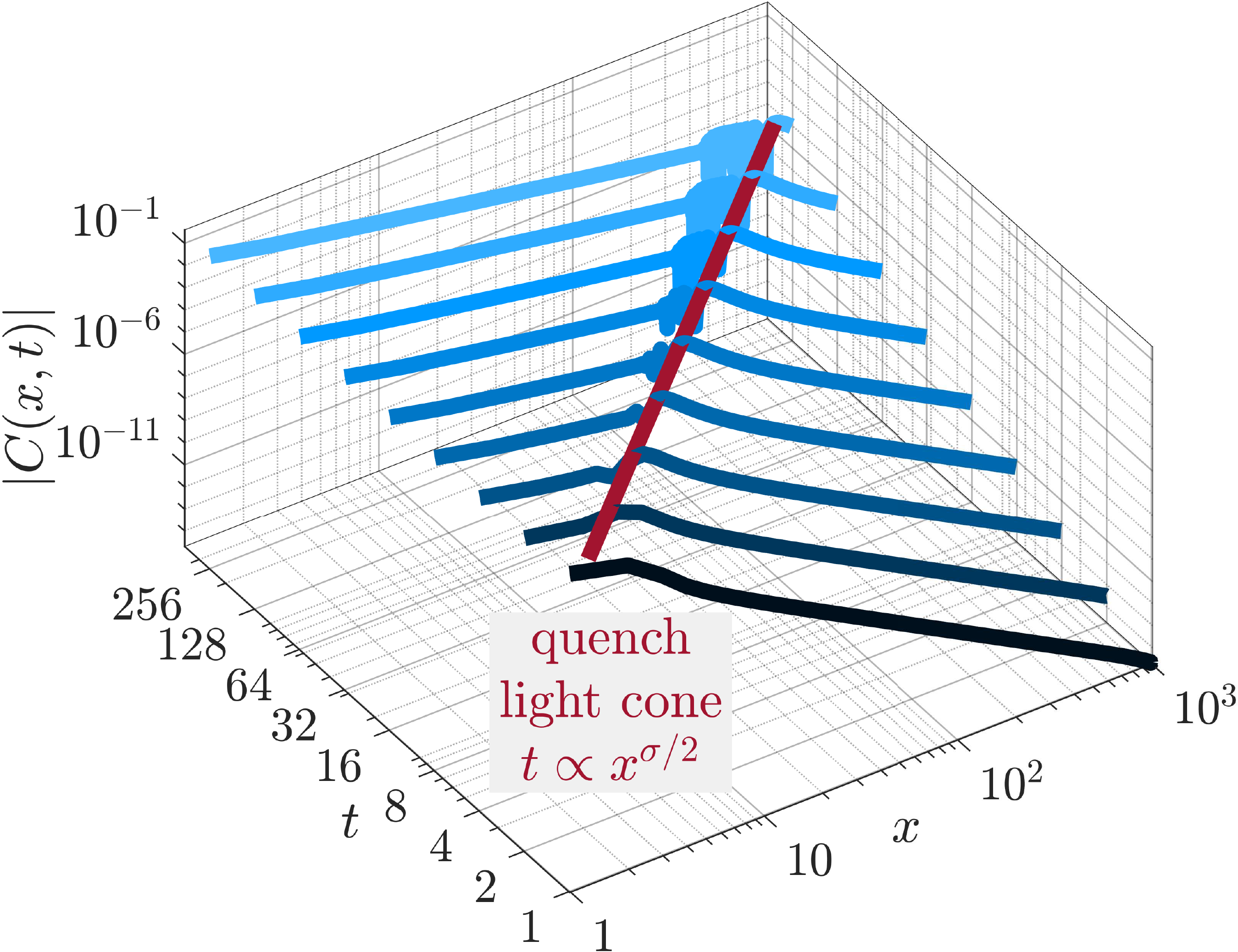}\\
\begin{minipage}[c]{0.48\linewidth}
\includegraphics[width=\linewidth]{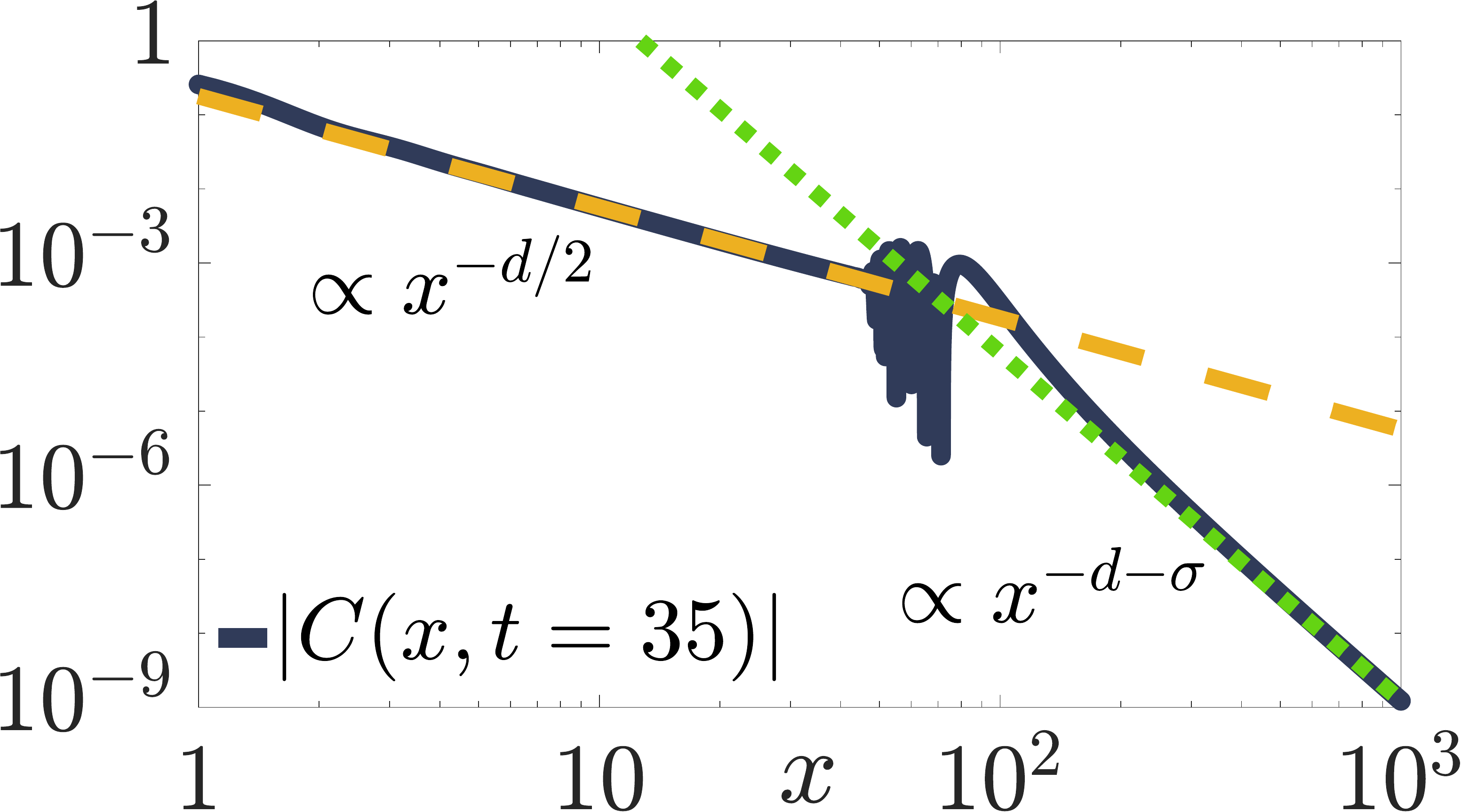}
\end{minipage}
\hfill
\begin{minipage}[c]{0.50\linewidth}
\includegraphics[width=\linewidth]{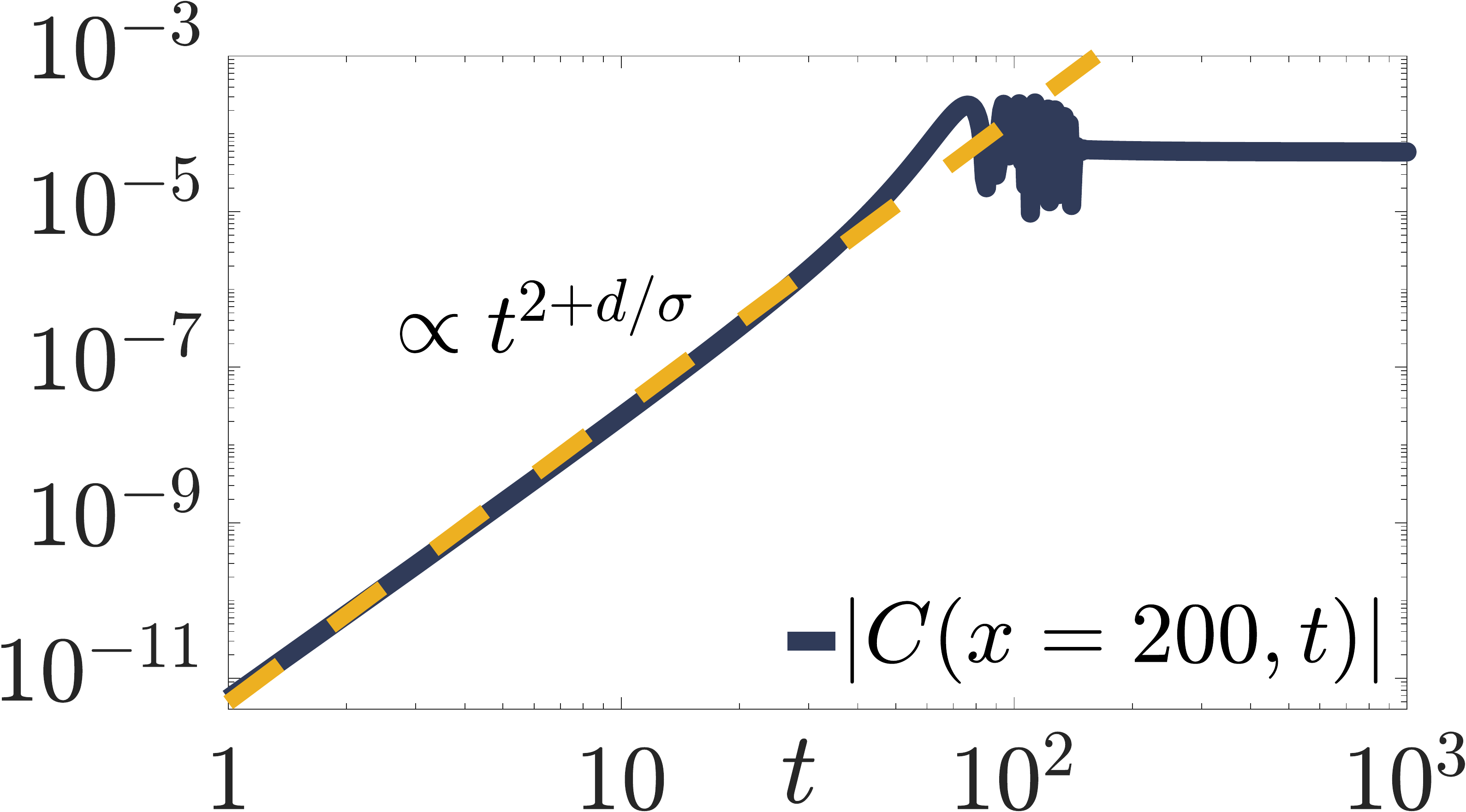}
\end{minipage}\caption{(Color online). Main panel: Equal-time correlation function as a function of time and distance in the wake of a critical quench ($r=r_c$); here, $\sigma=1.7$ and $d=3$. A nonlinear quench light cone emerges with $t\propto x^{0.85}$. Bottom left panel: Correlation function at fixed $t=35$. This function exhibits a crossover from the critical power law inside the quench light cone to $1/x^{d+\sigma}$ outside of it. Bottom right panel: Correlation function at a fixed distance $x=200$. This function grows as a power law in time before saturating to a constant value inside the quench light cone.}
	\label{fig:3D}
\end{figure}

Now we consider the correlation function in the stationary state well within the quench light cone; see \Fig{fig:summary}(a).  
To identify the nature of the stationary state, one must consider the long-time limit of the equal-time correlation function, which is given by \cref{eq:C-stationary}. Together with the value of $\alpha$ in \cref{Eq. alpha}, we find
\begin{align}\label{Eq. C_st}
    C_\st(\bk)\propto
    \frac{1}{k^{d/2}},  \qquad \sigma<d<2\sigma.
\end{align}
In the spatial domain, this equation implies that, well within the quench light cone, $x^{\sigma/2} \ll t$,  the correlation function falls off as
\begin{equation}\label{Eq. C_st}
    C_\st(\br)\propto \frac{1}{x^{d/2}}, \qquad \sigma<d< 2\sigma.
\end{equation}
Above the upper critical dimension, $d>2\sigma$, where the fixed point becomes Gaussian, we find $C_\st(\bk)\sim 1/k^\sigma$, which, in the spatial domain, gives $C_{\rm st}(\bx)\sim 1/x^{d-\sigma}$.

At intermediate times, \cref{Eq. C} hints at a crossover from the critical behavior inside the quench light cone $x\ll t^{2/\sigma}$ to a faster power-law decay due to the long-range coupling ($1/x^{d+\alpha}$)  outside the light cone $x\gg t^{2/\sigma}$. Putting Eqs.~(\ref{Eq. C}-\ref{eq:C-short-t}) together with \cref{Eq. alpha}, we have
\begin{align}\label{eq:crossover}
    C(\br,t)\propto
    \begin{cases}
    1/x^{d/2}, &  x \ll t^{2/\sigma}, \\
    t^{2+{d}/{\sigma}}/x^{d+\sigma}, & x \gg t^{2/\sigma},
    \end{cases}
\end{align}
at the non-Gaussian fixed point, $\sigma<d<2\sigma$. Interestingly, the correlation function increases algebraically in time outside the quench light cone. Above the upper critical dimension, $d>2\sigma$, we find  similar behavior where the power law $1/x^{d-\sigma}$ inside the quench light cone crosses over to $t^4/x^{d+\sigma}$ outside of it.

\begin{figure}[t]
	\centering
	\hspace{-.25 cm}\vspace{0.13cm}
	\includegraphics[width=.49\textwidth]{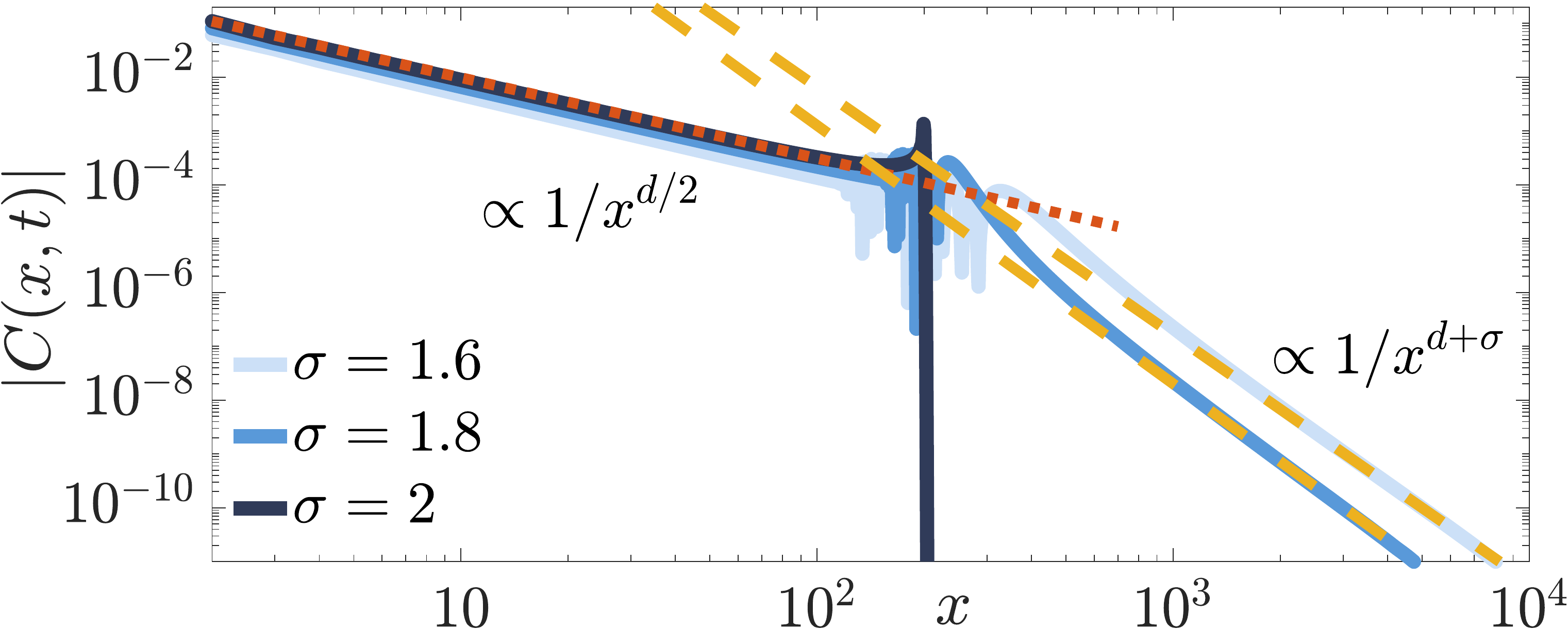}\\
	\caption{(Color online). Equal-time correlation function as a function of distance in the wake of a critical quench $r=r_c$; here, $t=100$ and $d=3$.
	For short-range coupling, $\sigma=2$, the correlation function decays exponentially.}
	\label{fig:Corr-critical}
\end{figure}

In \Fig{fig:3D}, we numerically show that indeed two distinct spatial power laws emerge inside and outside the quench light cone; see also the bottom left panel there. Furthermore, one can see that the correlation function increases as a power law with time outside the light cone consistent with the above equation; see the bottom right panel in \Fig{fig:3D}. These characteristics are schematically shown
in \Fig{fig:summary}(a).
Same qualitative features arise for different values of $\sigma$ as we show in \Fig{fig:Corr-critical}. Note that in the absence of the long-range coupling (when $\sigma=2$), correlations decay exponentially outside the light cone; see the thick curve in~\Fig{fig:Corr-critical}. 

There are a few interesting facts to note here.
The power-law decay of the correlation function [Eq.~\eqref{Eq. C_st}] at long times is independent of $\sigma$ at the non-Gaussian fixed point ($\sigma<d<2\sigma$). This is quite surprising given the long-range coupling of the model. This behavior is also characteristically different from the critical behavior of the long-range model in equilibrium either at zero or at finite temperature:
    \begin{equation}\label{Eq. C_eq}
        C_\eq(\br)\propto
        \begin{cases}
        1/x^{d-\sigma/2}, & T_c=0, \\
        1/x^{d-\sigma}, &  T_c>0,
        \end{cases}
    \end{equation}
    where $T_c$ denotes the critical temperature.
On the other hand, the power-law decay of the correlation function above the upper critical dimension depends on $\sigma$ in a fashion that is compatible with the equilibrium behavior at finite temperature; see Eq.~\eqref{Eq. C_eq}. This will be consistent with our analysis in \Sec{sec:eff-temp} where we argue that a finite effective temperature emerges above the upper critical dimension. On the other hand, we find that interestingly the effective temperature vanishes at long wavelengths as $\bk\to 0$ for $\sigma<d<2\sigma$.

\subsection{Response function}\label{sec:response-fn}
We have argued that, at long times, the response function becomes translation invariant [see \cref{Eq. G^R infty}] and recovers its form in equilibrium; the latter is identical to the response function of a quadratic Hamiltonian, and is thus independent of the state or the temperature of the system. 
In this subsection, we translate this statement into position space and derive scaling relations for the response function outside the local light cone. Most importantly, we argue that as long as $t-t' \ll t,t'$, the response function approaches its equilibrium value at \textit{any} distance $x$ inside, on, or outside either the quench or the causal light cone. To this end, let us assume $t-t' \ll t,t'$, and consider different spatial regions. We start with the region within the quench light cone, $x \ll t^{2/\sigma}$. This is exactly the domain of validity of \cref{Eq. G^R infty}, hence the stationary form of the response function. Next, we consider $x\gg (t-t')^{2/\sigma}$ well outside the local light cone and  resort to \cref{eq:GR-expanded}. 

\begin{figure}[t]
	\centering
	\hspace{-.25 cm}\vspace{0.13cm}
	\includegraphics[width=.46\textwidth]{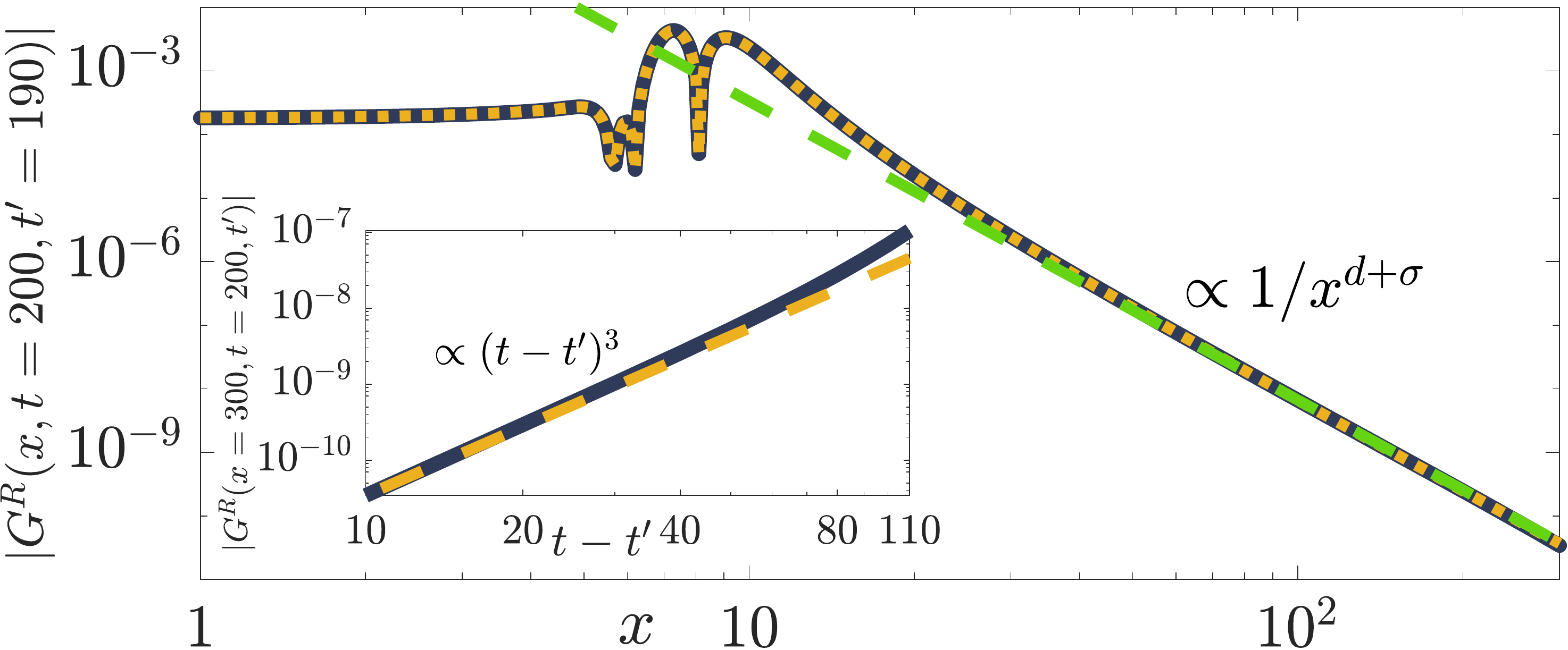}\\
	\caption{(Color online). Response function (solid dark blue curve) at fixed $t$ and $t'$ with $t-t'\ll t,t'$ as a function of distance in the wake of a critical quench ($r=r_c$) for $d=3$ and $\sigma=1.7$. The ``equilibrium'' response function in \cref{Eq. G^R infty-position} is an excellent fit (the dotted yellow curve). The response function outside the local light cone falls off as $1/x^{d+\sigma}$ (the dashed green line). Inset: Response function well outside the local light cone scales as $(t-t')^3$ at short time differences.}
	\label{fig:GR-eq}
\end{figure}

The latter equation is valid irrespective of the value of $k^{\sigma/2}t, k^{\sigma/2}t$ as long as $k^{\sigma}(t-t')\ll 1$. As noted in our discussion there,  \cref{eq:GR-expanded} is consistent with the equilibrium form of the response function in \cref{Eq. G^R infty}. It follows from these considerations that, for $t-t' \ll t,t'$,
at \textit{any} distance $x$,
\begin{equation}\label{Eq. G^R infty-position}
    G^R(\bx,t,t')=-\Theta(t-t')\int_\bk \frac{e^{i\bk\cdot\bx}}{k^{\sigma/2}}\sin\big[k^{\sigma/2}(t-t')\big].
\end{equation}
It is quite remarkable that, for short time differences, the response function \textit{equilibrates} globally at all distances; see \Fig{fig:summary}(b). 
In \Fig{fig:GR-eq}, we numerically verify this point and find an excellent agreement with the above equation.

We are particularly interested in the response function outside the local light cone, $x\gg (t-t')^{2/\sigma}$. This is obtained by expanding the sine function in the above equation to the third order where the first nonanalytical term ($\sim k^{\sigma}$) appears; see \cref{eq:GR-expanded}. We then find, for $t-t'\ll x^{\sigma/2},t$,
\begin{equation}
    G^R(\bx, t,t')\propto \frac{(t-t')^3}{x^{d+\sigma}}.
\end{equation}
Again, we numerically verify this scaling behavior in \Fig{fig:GR-eq} and the inset there.

Next we turn our attention to the response function in the limit $t'\ll t$. This regime exhibits aging and is discussed in more detail in the next subsection. 

\subsection{Aging}
\begin{figure}[t]
      \includegraphics[width=.23\textwidth]{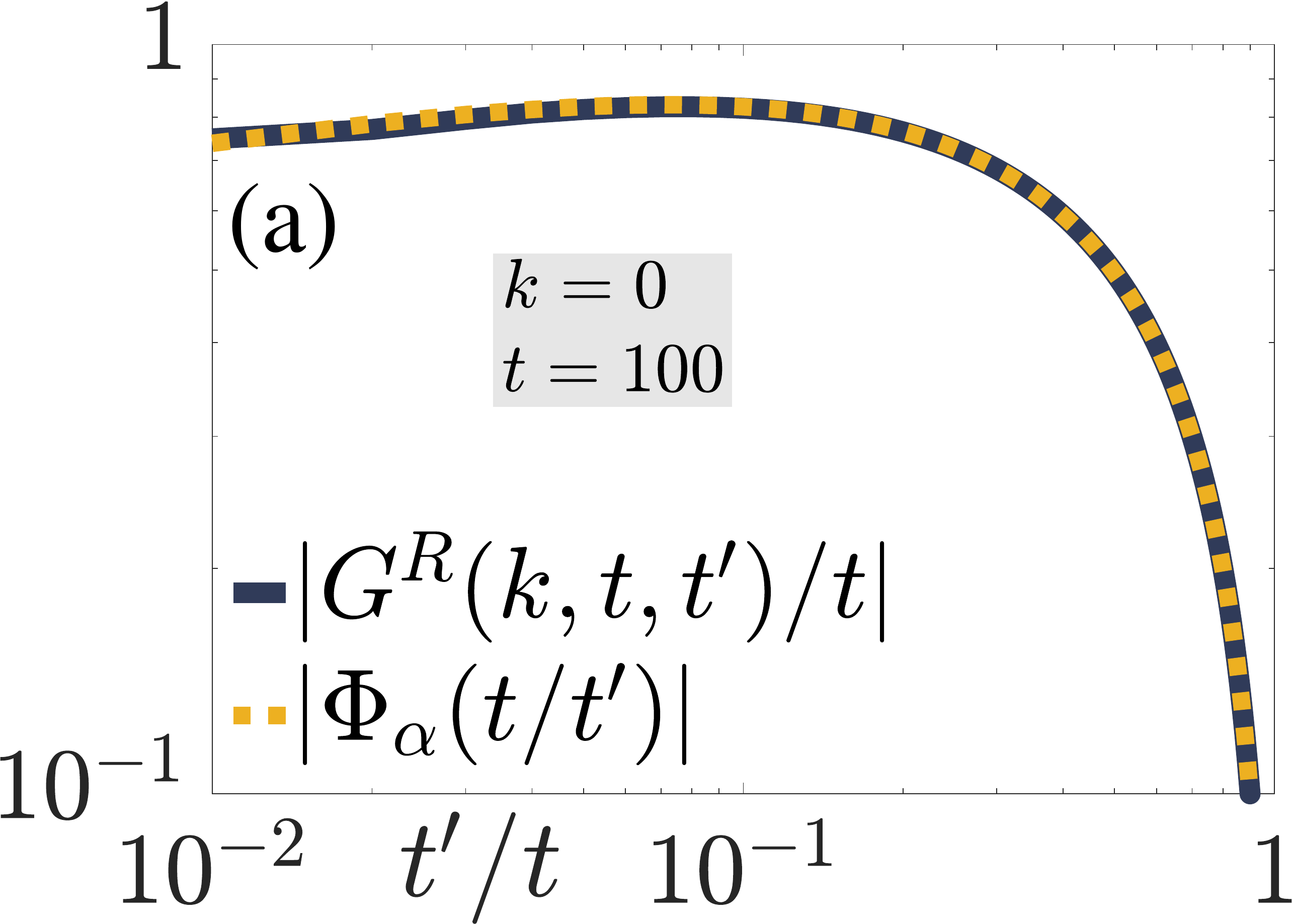}
      \includegraphics[width=.23\textwidth]{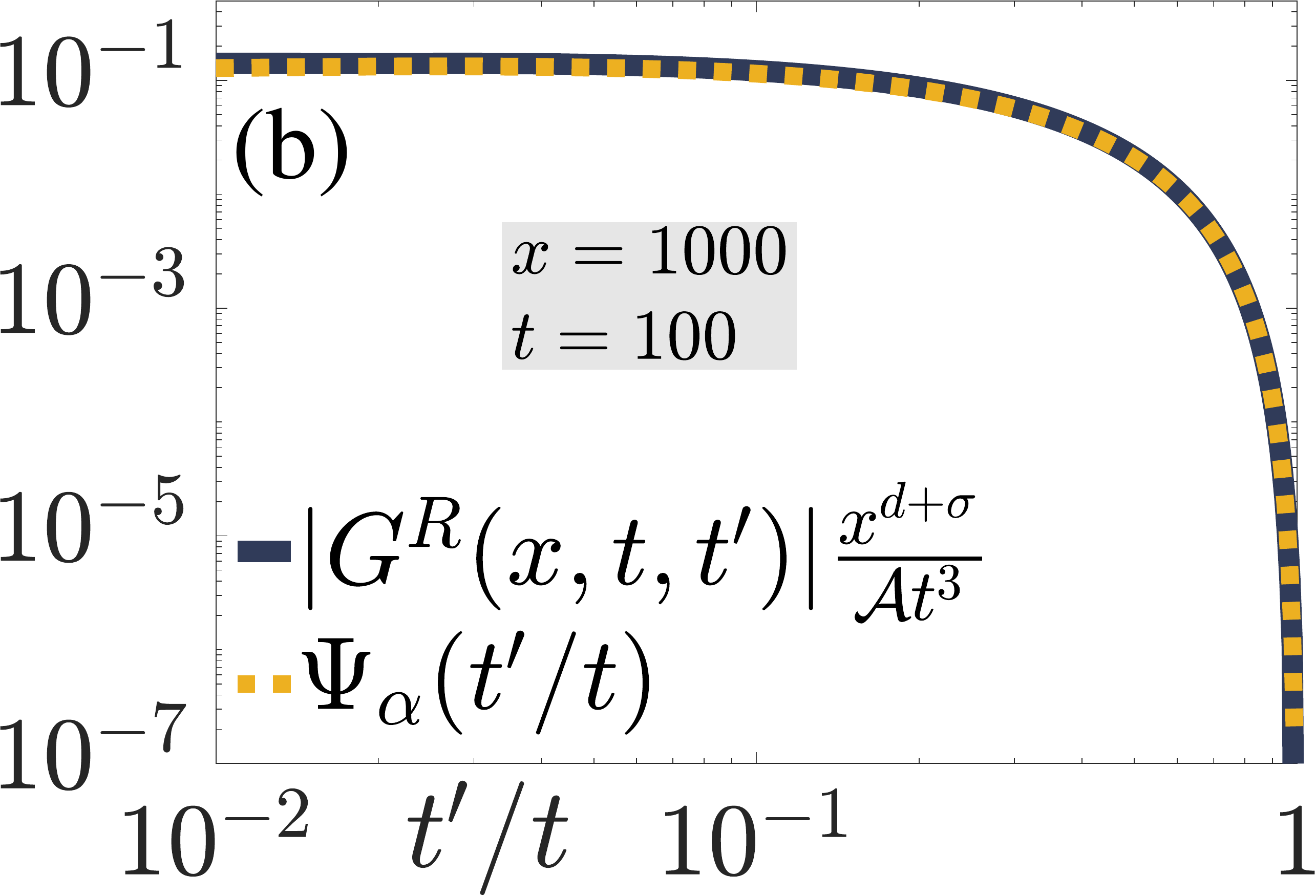}
    \caption{(Color online). Response function in the wake of a critical quench ($r=r_c$); here, $\sigma=1.7$ and $d=3$ and the larger time coordinate is set to $t=100$. (a) Response function in momentum space. The function $\Phi_\alpha(t/t')$ is an excellent fit. (b) Response function in position space at long distances well outside the quench light cone. The function $\Psi_\alpha(t'/t)$ is an excellent fit.}
    \label{fig_GR_critical}
\end{figure}

According to \cref{eq:GR-outside-light-cone}, the response function is proportional to $(t'/t)^\theta$ for $t'\ll t$. The initial-slip exponent $\theta$ is given by \cref{eq:th}, together with \cref{Eq. alpha}, as
\begin{equation}\label{eq:th-1}
    \theta=\frac{4-\tilde d}{4}=1-\frac{d}{2\sigma}, \qquad \sigma<d<2\sigma,
\end{equation}
while $\theta=0$ above the upper critical dimension, $d>2\sigma$. 
For arbitrary $t'/t$ and at $\bk\to 0$, the response function is given by [see \cref{eq:GR-outside-aging1}]
\begin{equation}\label{eq:GR-Phi}
    G^R(\bk\to 0,t,t')\sim t \, \Phi_\alpha(t'/t),
\end{equation}
with the function $\Phi$ defined in \cref{eq:Phi} and $\alpha=(d/\sigma-1)/2$ as follows from \cref{Eq. alpha}; note that $\Phi_\alpha(y)\propto y^\theta$ for small $y$. We find an excellent agreement between the above analytical expression and the numerical simulation in \Fig{fig_GR_critical}(a).

Next, we consider the response function in position space. To describe the dependence on $t'/t$, we first point out that the response function is proportional to $t'^\theta$ as long as $t'\ll t$. This is because, in this limit, a natural cutoff for the momentum integral is roughly set by $k\lesssim t^{-2/\sigma}$. Now the domain of the Bessel function $J_{\pm\alpha}(k^{\sigma/2}t')$ is of the order $\lesssim t'/t$, therefore it can be expanded to the first nontrivial order to give $G^R \sim t'^\theta$. In the following, however, we focus on the response function outside the quench light cone, $x\gtrsim t^{2/\sigma}$; see \Fig{fig:summary}(c). From \cref{eq:GR-outside-aging1}, we find that, at long distances $x^{\sigma/2}\gtrsim t,t'$, 
\begin{equation}\label{eq:GR-Psi}
    G^R(\bx,t,t')\sim {\cal A} \frac{t^3}{x^{d+\sigma}} \, \Psi_\alpha(t'/t),
\end{equation}
where the function $\Psi$ is defined in \cref{eq:Psi} together with $\alpha=(d/\sigma-1)/2$; note that $\Psi(y)\propto y^\theta$. The constant of proportionality $\cal A$ in this equation is simply given by the inverse Fourier transform of $k^{\sigma}$ in position space; for $d=3$, for example, we have ${\cal A}=\Gamma(2+\sigma)\sin(\pi\sigma/2)/2\pi^2$. We plot the response function in \Fig{fig_GR_critical}(b) well outside the quench light cone and find excellent agreement with the analytical expression above. 

Finally, we point out that the classical (stochastic) $O(N\to\infty)$ model leads to the same initial-slip exponent $\theta$ given by \cref{eq:th-1} although the overall dependence on $t$ is different; see Appendix \ref{app:classical}. The same exponent governs the long-range spherical model \cite{Henkel11}. We also remark that the same conclusion emerges in the short-range variant of the quantum model considered here \cite{Maraga2015,Chiocchetta2016b}.

\subsection{Effective temperature}\label{sec:eff-temp}
In equilibrium, correlation and response functions are related by the celebrated fluctuation-dissipation relation as (suppressing the $\bk$ dependence for now) \cite{kamenev_book}
\begin{equation}
    G^K(\omega)=\hbar \coth\left(\frac{\hbar\omega}{2T}\right) \left[ G^R(\omega)-G^R(-\omega)\right].
\end{equation}
In the high-temperature limit ($T\gg \hbar \omega$), one recovers the \textit{classical} limit of the fluctuation-dissipation relation, $G^K(\omega)=(2T/\omega)  \left[G^R(\omega)-G^R(-\omega)\right]$, or, in the time domain, $i\partial_t G^K(t-t')= 2T\left[ G^R(t-t')-G^R(t'-t)\right]$. For $t>t'$, we simply have $i\partial_t G^K(t-t')= 2T G^R(t-t')$.

Even outside equilibrium, the fluctuation-dissipation relation can be used as a basis to define an effective temperature \cite{Cugliandolo11} which may nevertheless depend on $\bk$.
To this end, we must characterize the time dependence of the correlation and response functions. One can easily deduce the unequal-time correlation function at long times ($|t-t'|\ll t,t'$) as
\begin{equation}
   iG^K(\bk,t,t')=C_\st(\bk)\cos\left[k^{\sigma/2} (t-t')\right].
\end{equation}
We remind the reader that $C_\st(\bk)$ denotes the equal-time correlation function in the stationary state.
Now, together with Eq.~\eqref{Eq. G^R infty}, it becomes clear that the correlation and response functions satisfy the \textit{classical} form of the fluctuation-dissipation relation
\begin{equation}
    i\partial_t G^K(\bk, t-t')=2 T_\eff(\bk) G^R(\bk,t-t'),
\end{equation}
for a $\bk$-dependent effective temperature
\begin{equation}
    T_\eff(\bk)\propto k^{\sigma-d/2}, \qquad \sigma<d<2\sigma.
\end{equation}
In contrast, we find that $T_\eff =$ const, i.e., independent of $\bk$, at the Gaussian fixed point above the upper critical dimension, $d>2\sigma$. The scaling of the effective temperature with momentum is intriguing and has important consequences. Most importantly, it indicates that the system is genuinely out of equilibrium as the critical, long-distance properties are not captured by a constant effective temperature below the upper critical dimension. In fact, the exponent governing the scaling of the effective temperature is simply $\sigma \theta$, that is, the critical exponet describing the effective temperature at late times is directly proportional to the aging exponent characterizing the short-time dynamics. This fact is indicative of the true nonequilibrium nature of the stationary state \cite{Maraga2015,Chiocchetta2016b}.
Furthermore, the effective temperature vanishes with $\bk \to 0$, which might seem to indicate that quantum fluctuations become important; this would in turn invalidate the classical limit of effective temperature. However, the classical description is still valid if the frequency $\omega_\bk\sim k^{\sigma/2}$ of a given mode $\bk$ is small compared to the temperature, $T_\eff(\bk)$. A quick comparison shows that this is always the case provided that $d>\sigma$ that is trivially satisfied above the lower critical dimension. Finally, the fact that the effective temperature is simply a constant above the upper critical dimension is compatible with a simple analysis of harmonic oscillators appropriate to the corresponding Gaussian fixed point \cite{ciocchetta_2015,Maraga2015,Chiocchetta2016b}.

\section{Quench below the dynamical critical point $r<r_c$}\label{sec:subcritical}
Despite many parallels with the previous section, we find it more appropriate to devote another section to the quench below the dynamical critical point for a clearer presentation and to highlight the contrasts with the critical quench. 

In equilibrium and below the dynamical critical point, the systems becomes ordered,
\begin{equation}
	\lim_{|\bx|\to \infty}C_\eq(\br)= {\rm const} >0,
\end{equation}
or, alternatively in momentum space, $C_\eq(\bk)\sim V \delta(\bk)$ with $V$ the system's volume. The dynamics of phase ordering is often described by expanding domains, in a process known as coarsening \cite{Bray02}. However, the quantum quench considered here does not give rise to ordering even for $r<r_c$ as we shall see below. Instead, it will give rise to new critical behavior, distinct from that in a quench to the dynamical critical point. 
Such behavior can also be viewed as an anomalous coarsening dynamics \cite{Chandran13,Sciolla2013,Maraga2015}.
With no upper critical dimension, the nontrivial (i.e., non-Gaussian) critical behavior in this case persists at any dimension $d>d_l$. In this section, we derive analytical results for the correlation and response functions and aging behavior and compare them against exact numerical results. Finally, we identify a $k$-dependent effective temperature that captures the critical correlations.

\subsection{Correlation function}
To identify the nature of the stationary state, we resort to \cref{eq:C-stationary}. For a quench below the dynamical critical point, we must substitute $\alpha$ by the value given in \cref{Eq. alpha'}:
\begin{align}\label{Eq. C_st-1}
    C_\st(\bk)\propto
    \frac{1}{k^{d-\sigma/2}},
\end{align}
for any dimension above the lower critical dimension $d>d_l=\sigma$. In the spatial domain, this means that the correlation function falls off as
\begin{equation}\label{Eq. C_st-2}
    C_\st(\br)\propto \frac{1}{x^{\sigma/2}}.
\end{equation}
Interestingly, the decay of the correlation function is independent of dimensionality. This becomes particularly surprising given that, in higher dimensions, a slow decay of correlations indicates a highly correlated system. 
Indeed, we shall see that, at sufficiently high dimensions, the effective temperature in the stationary state can even diverge as $\bk\to 0$.
\begin{figure}[t]
	\centering
	\hspace{-.25 cm}\vspace{0.13cm}
	\includegraphics[width=.49\textwidth]{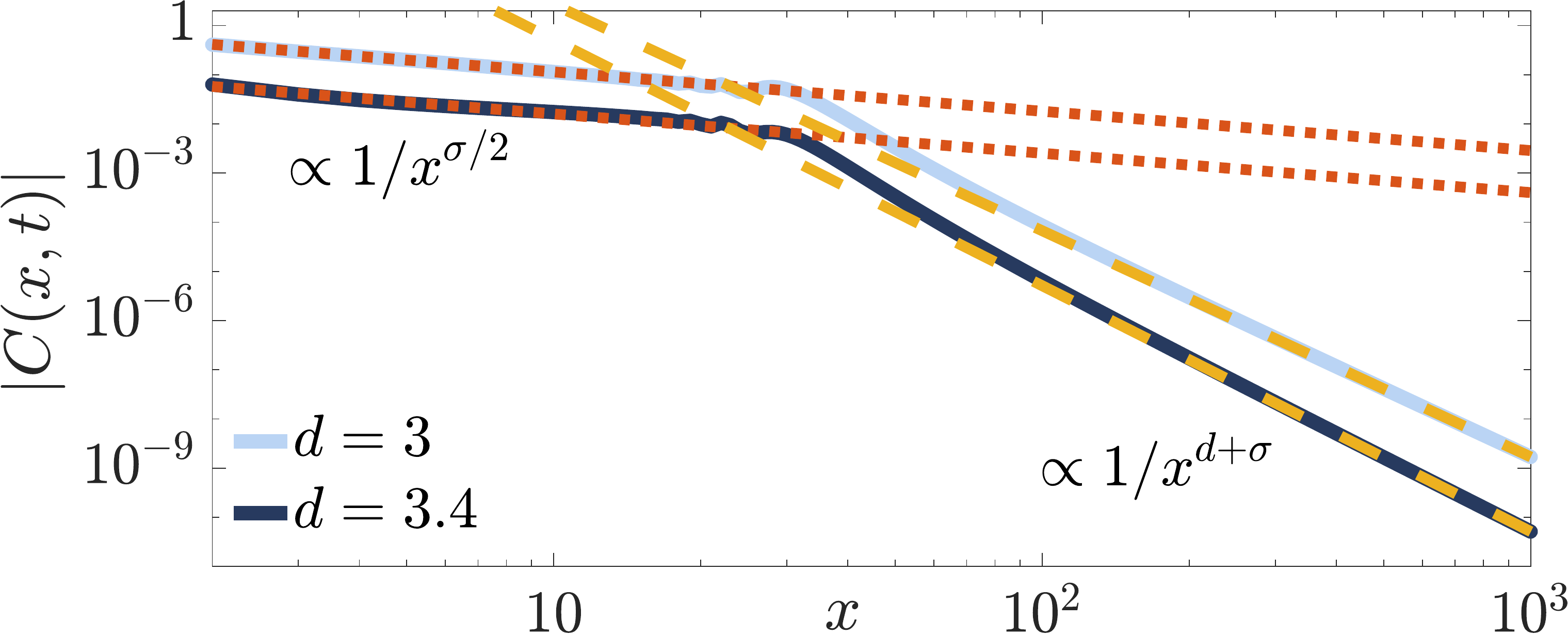}\\
	\caption{(Color online). Equal-time correlation function (solid blue curves) as a function of distance in the wake of a quench below the dynamical critical point ($r=3r_c$); here, $t=15$ and $\sigma=1.6$.}
	\label{fig:fig5}
\end{figure}

In this case too, we expect a crossover at a time scale governed by the nonlinear light cone, $x\propto t^{2/\sigma}$. 
Putting Eqs.~(\ref{Eq. C}-\ref{eq:C-short-t}) together with \cref{Eq. alpha'}, we have 
\begin{align}
    C(\br,t)\propto
    \begin{cases}
    1/x^{\sigma/2}, & x \ll t^{2/\sigma}, \\
    t^{1+2d/\sigma}/x^{d+\sigma}, & x \gg t^{2/\sigma}.
    \end{cases}
\end{align}
A stationary, time-independent power-law correlation function ($1/x^{\sigma/2}$) within the nonlinear quench light cone crosses over to the long-range power law ($1/x^{d+\sigma}$) outside the light cone. 
In \Fig{fig:fig5}, we verify the emergence of these distinct power laws for different values of $\sigma$.

\subsection{Response function \& aging}	
An almost identical analysis to \Sec{sec:response-fn} shows that the response function ``equilibrates'' globally at all distances when $t,t'\gg t-t'$ and is identical to that in a critical quadratic Hamiltonian, hence no dependence on the state; cf. \Fig{fig:GR-eq}.

Furthermore, the analysis of aging behavior closely follows our discussion in the previous section. Only for the quench below the dynamical critical point, $r<r_c$, we find a distinct initial-slip exponent 
\begin{equation}\label{eq:th-2}
	\theta=\frac{4-\tilde d}{2}=\frac{3}{2}-\frac{d}{\sigma}, \qquad \sigma<d.
\end{equation}
Interestingly, the exponent $\theta$ becomes negative for $d>3\sigma/2$. As  we shall discuss in the next subsection, the effective temperature diverges as $\bk\to 0$ in this regime.

\begin{figure}[t]
	\centering
	\hspace{-.25 cm}\vspace{0.13cm}
	\includegraphics[width=.45\textwidth]{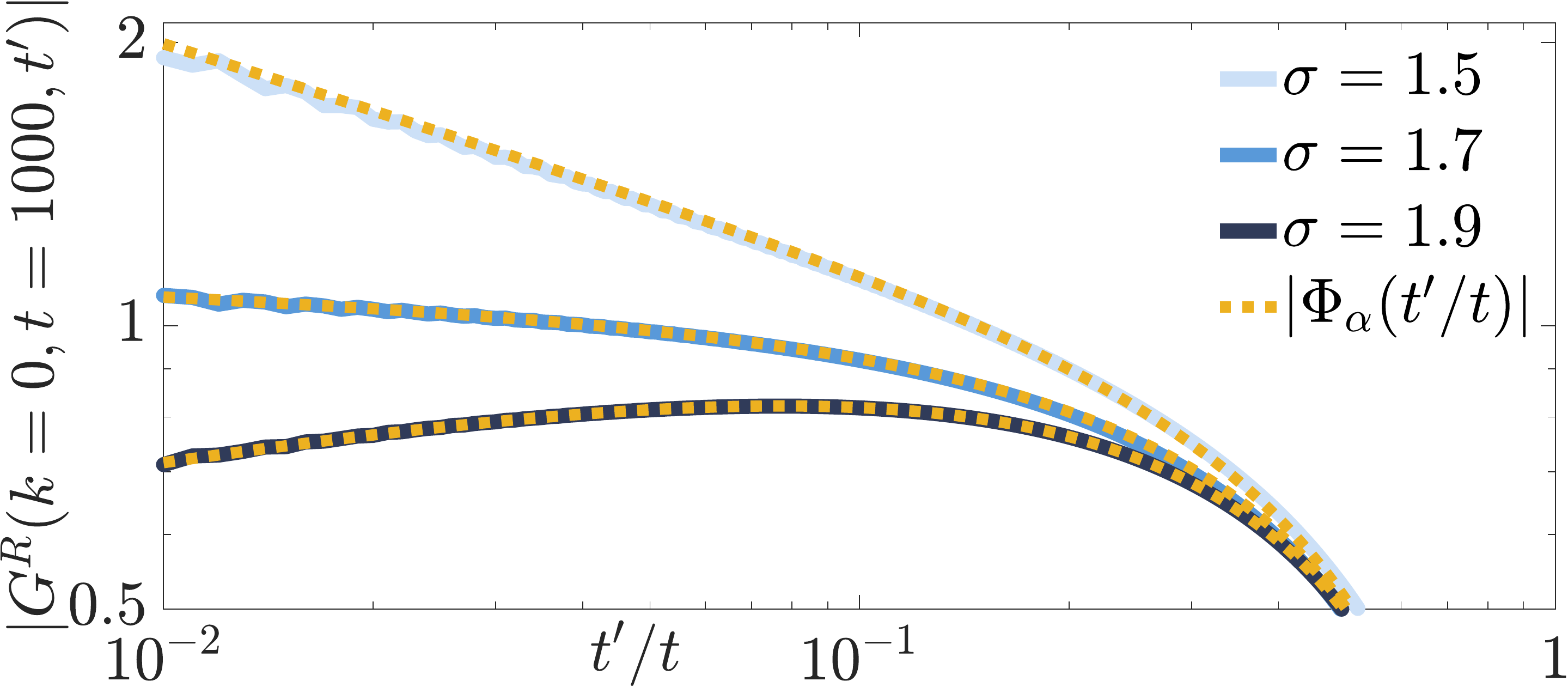}\\
	\caption{(Color online). Response function (solid blue curves) in the wake of a quench below the dynamical critical point, $r=3r_c$; here, $d=2.6$ and the larger time coordinate is set to $t=1000$. The plots are consistent with the response function increasing as $t'/t$ decreases for $\sigma<2d/3$.}
	\label{fig:fig6}
\end{figure}

Again, we can identify aging behavior by probing the dependence on $t'/t$ either in momentum space via \cref{eq:GR-Phi} or in position space, outside the quench light cone, via \cref{eq:GR-Psi} but with the value of $\alpha=d/\sigma -1$ from \cref{Eq. alpha'}. Figure \ref{fig:fig6} depicts the dependence of the response function on $t'/t$ in momentum space in the limit $\bk\to0$. In this figure, one can clearly see that the response functions corresponding to negative values of $\theta$ further increase as the ratio $t'/t$ becomes smaller at fixed $t$; this is in contrast with the behavior for $\theta>0$; see also~\Fig{fig_GR_critical}(a).

\subsection{Effective temperature}\label{Sec. Effective temperature2}
An effective temperature that captures the critical fluctuations can be identified similar to \Sec{sec:eff-temp}. Similar line of reasoning, based on the fluctuation-dissipation relation, leads to
\begin{equation}
T_\eff(\bk)\propto k^{3\sigma/2-d},
\end{equation}
for a quench below the dynamical critical point, $r<r_c$.
As remarked earlier, this effective temperature can even diverge with $\bk \to0$ depending on dimensionality $d$ and the exponent $\sigma$ when $d>3\sigma/2$. It then follows that, in this regime, the density of low-momentum states diverges faster than that at any finite temperature. Again, we notice that the exponent in the above equation is simply $\sigma \theta$, underscoring the nonequilibrium nature of the stationary state. The effective temperature thus diverges at long wavelengths precisely in a regime where the initial-slip exponent becomes negative.

\section{Conclusion}\label{sec:conc}
We have considered quenches to and below the dynamical critical point of the $d$-dimensional quantum $O(N\to\infty)$ model with the long-range coupling $1/x^{d+\sigma}$, and have investigated the critical behavior of the ensuing short-time dynamics as well as the sationary state at late times through the two-point correlation and response functions. Qualitative behavior of the two-point functions depends on the corresponding space-time positions with respect to both local and quench light cones. While the correlation function saturates to a stationary value inside the quench light cone and exhibits critical scaling distinct from that in equilibrium, it grows algebraically with time outside this light cone while falling off with the distance as $1/x^{d+\sigma}$, the latter in harmony with the Hastings-Koma bound \cite{Hastings05}. The response function too becomes stationary as long as the time difference $t-t'$ is short compared to the age of the quenched quantum system, and becomes identical to that in equilibrium. Aging becomes manifest in the limit $t' \ll t$ where the response function scales as $(t'/t)^\theta$ with the initial-slip exponent $\theta$ that characterizes a truly out-of-equilibrium critical behavior. This exponent along with other scaling properties exhibit a rich interplay between dimensionality and long-range coupling, and are distinct for quenches to or below the dynamical critical point. Indeed, a surprising observation is that even a quench below the dynamical critical point exhibits critical scaling, or an anomalous type of coarsening \cite{Chandran13,Smacchia2015,Maraga2015}, in contrast with the phase ordering and coarsening dynamics that would occur when the system is coupled to a thermal bath  \cite{Bray02}. 

Further investigation of the quantum quench well below the dynamical critical point, the role of the corresponding initial-slip exponent and its connection with the order-parameter dynamics is worthwhile. Another direction is to investigate consequences of integrability breaking, for example due to finite (but large) $N$. Crossover to thermalization at long times will showcase the nontrivial dynamics that is constrained by a weakly broken integrability. 
Finally, it would be interesting to identify qualitative features of the quench dynamics of the $O(N)$ model that extend to other systems such as spin models with long-range interactions.

%\section*{Acknowledgments}
\begin{center}
    \small\textbf{ACKNOWLEDGMENTS}
\end{center}

The authors acknowledge stimulating discussions with Cameron Kilgore, Jamir Marino, and Francesco Piazza, and are grateful to Alessio Chiocchetta for a critical reading of the manuscript and for valuable comments. J.C.H. acknowledges support by the DFG Collaborative Research Centre SFB 1225 (ISO- QUANT) and the ERC Starting Grant StrEnQTh. M.M. acknowledges support from NSF under Grant No. DMR-1912799 and start-up funding from Michigan State University. 

\appendix

 \section{Equilibrium and dynamical critical points}\label{sec:CriticalPoint}
The long-range quantum $O(N\to\infty)$ model given by Eq.~\eqref{eq:Ham} can be described in the disordered phase, where $\langle\Phi\rangle=0$, by
\begin{align}
r _\text{eff}= r+\frac{u}{12}\int_k\frac{1}{\sqrt{k^\sigma+r_\text{eff}}}h(k/\Lambda).
\end{align}
The equilibrium critical point is then readily calculated by setting $r _\text{eff}=0$, yielding
\begin{align}
	r_{\text{eq},c}=-\frac{u}{12}\int_kk^{-\sigma/2}h(k/\Lambda).
\end{align}

On the other hand, in order to determine the dynamical critical point of the long-range quantum $O(N\to\infty)$ model and following Refs.~\cite{Sotiriadis10,Smacchia2015,Maraga2015}, we make the ansatz that, at late times, the equal-time correlation function is the same as the stationary part of the free (noninteracting) theory with the initial parameter $r_0=\Omega_0^2$ and the final parameter $r_f$ to be self-consistently determined. With this ansatz, we just replace $r_{\rm eff}$ in \cref{Eq. r_eff-1} with $r_f$:
\begin{align}\label{eq:diffeq}
\ddot{f}_\bk+\left[k^\sigma+r_f\right]f_\bk=0,
\end{align}
again with initial conditions $f_\bk(0)=1/\sqrt{2\omega_{0\bk}},\,\dot{f}_\bk(0)=-i\sqrt{\omega_{0\bk}/2}$ where $\omega_{0\bk}=\sqrt{k^\sigma+r_0}$. Inserting the late-time solution in \cref{Eq. r_eff-2} and keeping the stationary (non-oscillatory) contribution, we find 
\begin{equation}
    \lim_{t\to \infty}r_{\rm eff}(t)=r_f+ \frac{u}{4!}\int_\bk \frac{2k^\sigma+\Omega_0^2}{k^\sigma\sqrt{k^\sigma+\Omega_0^2}}h(k/\Lambda).
\end{equation}
We can then infer the dynamical critical point as
\begin{align}\label{eq:CP}
r_c=&\,-\frac{u}{4!}\int_\bk \frac{2k^\sigma+\Omega_0^2}{k^\sigma\sqrt{k^\sigma+\Omega_0^2}}h(k/\Lambda).
\end{align}
The approximate ansatz used here provides accurate predictions for the dynamical critical point~\eqref{eq:CP},  according to our numerical simulations. Notice that as the dimension approaches the lower critical dimension, $d_l=\sigma$, we have $r_c\to -\infty$ since the integral diverges due to the contribution near $k= 0$, hence the absence of ordering at $d=d_l$.

\section{Classical aging with long-range interactions}\label{app:classical}
In this section, we present the classical aging behavior in the presence of long-range interactions \cite{Chen2000,Chen2001,Chen2002,Baumann2007,Corberi2019}. We closely follow the treatment by Jenssen \textit{et al}. for the short-range classical model \cite{Janssen1989} which naturally reduces to the short-range result upon inserting $\sigma=2$. Alternative derivation for the (long-range) spherical model is provided in Ref.~\cite{Henkel11}.
We warn the reader that the conventions used here for the response and correlation functions are different from those in the main text and reflect the standard notation in the classical literature.
We begin with the classical Hamiltonian 
\begin{equation}
    {\cal H}=\int \diff^d\bx \left[-\frac{1}{2}\bphi\cdot\nabla^\sigma\bphi+\frac{r}{2}\bphi^2+\frac{u}{4!N}(\bphi^2)^2\right],
\end{equation}
and the associated stochastic (Langevin) equation
\begin{equation}
    \partial_t \bphi=-\lambda \frac{\delta{\cal H}}{\delta \bphi}+\bxi(t),
\end{equation}
with the noise correlations
\begin{equation}
    \langle \xi_a(\bx,t) \xi_b(\bx', t')\rangle=2\lambda \delta_{ab}\delta(\bx-\bx')\delta(t-t').
\end{equation}
The parameter $\lambda$ can be viewed as \textit{mobility} and its appearance in both the equation of motion as well as the noise correlator reflects the fluctuation-dissipation relation. By introducing the response filed \cite{kamenev_book}, one can describe the stochastic dynamics in a functional-integral language as
\begin{equation}
    Z=\int \diff \bphi \diff \tilde\bphi\,e^{-J[\tilde\bphi,\bphi]},
\end{equation}
with the functional $J$ given by
\begin{align}
    J&[\tilde\bphi,\bphi]=\int_0^\infty \diff t \int \diff \bx  \nonumber \\
    &\left[\tilde \bphi\cdot\left(\dot \bphi+\lambda[-\nabla^\sigma+r]\bphi+\frac{\lambda u}{6N}\bphi \bphi^2\right) -\lambda \tilde\bphi^2\right]. 
\end{align}
Again, in the large-$N$ limit, we can replace the nonlinear term by 
\begin{equation}
    \frac{1}{N} (\tilde \bphi \cdot\bphi) \bphi^2  \longrightarrow   C(t) \tilde \bphi \cdot\bphi,
\end{equation}
where 
\begin{equation}
    C(t)\equiv \frac{1}{N} \langle \bphi^2(t)\rangle.
\end{equation}
With this substitution, exact in the limit $N\to \infty$, the functional $J$ becomes 
\begin{equation}\label{Eq. J Gauss}
    J[\tilde\bphi,\bphi]=\int_{t,\bx} 
    \left[\tilde \bphi\cdot\left(\dot \bphi+\lambda[-\nabla^\sigma+r_{\rm eff}(t)]\bphi\right) -\lambda \tilde\bphi^2\right],
\end{equation}
where the mass term is replaced by 
\begin{equation}\label{eq:app:self-consistent}
    r\longrightarrow  r_{\rm eff}(t)=r+ \frac{u}{6}C(t).
\end{equation}
The functional $J$ being quadratic, the ensuing dynamics can be solved exactly. Specifically, the correlation and response functions can be determined by inverting the kernel in the functional $J$; we find, in momentum space, 
\begin{align}
    \chi(\bk,t,t')&=\Theta(t-t') \exp\Big\{-\lambda \int_{t'}^t \diff s [r_{\rm eff}(s)+k^\sigma]\Big\}, \\
    C(\bk,t,t')&=2\lambda \int_0^\infty \diff s \,\chi(\bk,t,s)\chi(\bk,t',s), \label{Eq. App self-cons}
\end{align}
for the (retarded) response function $\chi$ and the correlation function $C$, respectively; notice that $C(t)=\int_\bk C(\bk, t,t)$. The self-consistent condition in \cref{eq:app:self-consistent} can be now stated as 
\begin{align}
    r_{\rm eff}(t)
    &=r+\frac{u}{6}\int_\bk C(\bk, t) \nonumber \\
    &=r_{\rm eff}(\infty) +\frac{u}{6}\int_\bk [C(\bk, t)-C(\bk, \infty)],
\end{align}
where we have introduced $C(\bk,t,t)\equiv C(\bk,t)$ for brevity. The integral over momentum is up to a cutoff momentum that we conveniently choose to be $\Lambda=1$. In a quench to the critical point, we have $r_{\rm eff}(\infty)=0$ and the correlation function at late times is determined by that of critical Gaussian model at finite temperature, i.e.,  $C(\bk, \infty)=1/k^\sigma$. 
Comparing various terms in \cref{Eq. J Gauss}, a scale-invariant solution emerges at intermediate times only if 
\begin{equation}
    r_{\rm eff}(t)=\frac{a}{2\lambda t},
\end{equation}
with $a$ a dimensionless constant. Notice that this effective mass scales with time in the same fashion as the first time derivative in \cref{Eq. J Gauss}. 
Inserting this equation in the response and correlation functions, we find
\begin{align}
\chi(\bk, t,t')&=\Theta(t-t')\left(\frac{t'}{t}\right)^{a/2} \exp\{-\lambda(t-t')k^\sigma\}, \\
C(\bk,t,t')&=2\lambda\int_0^{{\rm min}(t,t')} \hspace{-.5cm}\diff s\, \left(\frac{s^2}{tt'}\right) \exp\{-\lambda(t+t'-2s)k^\sigma\}.
\end{align}
It follows from the last equation that the equal-time correlation function at intermediate times can be written as
\begin{equation}
    C(\bk,t)=\frac{1}{k^\sigma}F(2\lambda k^\sigma t).
\end{equation}
Apart from the coefficient $2\lambda$ that is chosen for later convenience, the argument of the scaling function is dictated by the dynamical exponent.\footnote{The classical dynamical exponent $z=\sigma$ should be contrasted with $z=\sigma/2$ of the quantum model considered in the main text. For short-range coupling (upon inserting $\sigma=2$), the classical dynamical exponent $z=2$ characterizes diffusive dynamics in contrast with $z=1$ of the quantum model which underlies light-cone dynamics.} The precise form of the scaling function is given by
\begin{equation}\label{Eq. F}
    F(x)=x\int_0^1 \diff y\, (1-y)^a e^{-xy}.
\end{equation}
Incidentally, the scaling takes exactly the same form as that of short-range interactions \cite{Janssen1989}; this is evident in the fact that the function $F$ is independent of $\sigma$ which only enters through the definition of the argument of the scaling function ($\propto k^\sigma t$). 

Now in a quench to the critical point, $r_{\rm eff}(\infty)=0$, \cref{Eq. App self-cons} upon a change of variables, $z=2\lambda k^\sigma t$, yields
\begin{equation}
    a= \frac{u K_d }{6\sigma}(2\lambda t)^{2-d/\sigma}\int_0^{2\lambda t} \diff z\, z^{ d/\sigma-2}\left[F(z)-1\right].
\end{equation}
While the lhs is a constant, the rhs appears to increase with time. To ensure self-consistency, we must require that the leading time-dependent term vanishes, 
\begin{equation}
    \int_0^{\infty} \diff z\, z^{ d/\sigma-2}\left[F(z)-1\right]=0.
\end{equation}
Using \cref{Eq. F}, the latter condition amounts to 
\begin{equation}
    \frac{\Gamma(a+1)\Gamma(2-d/\sigma)}{\Gamma(a+2-d/\sigma)}=0,
\end{equation}
with the only consistent solution being
\begin{equation}
    a=-(2-d/\sigma).
\end{equation}
With this, we can now determine the response function as 
\begin{equation}\label{eq:chi-classical}
    \chi(\bk,t,t')=\Theta(t-t')\left(\frac{t}{t'}\right)^{1-\frac{d}{2\sigma}} \exp\{-\lambda(t-t')k^\sigma\},
\end{equation}
from which we can identify the aging exponent
\begin{equation}
    \theta=1-\frac{d}{2\sigma}.
\end{equation}
We remark that, similar to the quantum aging discussed in the main text, the aging exponent can be determined from the short-range case by replacing the dimension $d$ by $\tilde d=2d/\sigma$. The classical initial-slip exponent is identical to this exponent in our quantum model; see \cref{eq:th-1}. However, note that the response function of the quantum model in the limit $\bk \to0$ comes with an additional factor of $t$ \cite{ciocchetta_2015,Maraga2015,Chiocchetta2016b}. We can also investigate the response function at long distances [cf. \Fig{fig:summary}(c)]. It follows from \cref{eq:chi-classical} that, in a regime where $t' \ll t \ll x^\sigma$, 
\begin{equation}
    \chi(\bx,t,t')\propto \frac{t}{x^{d+\sigma}} (t'/t)^\theta.
\end{equation}
This is similar to \cref{eq:GR-Psi} in the limit $t'\ll t$ with the difference that there is an overall cubic, rather than linear, dependence on $t$ in the quantum model. 

Finally, we mention that a quench below the critical point leads to an ordered phase at late times and coarsening dynamics at intermediate times, in contrast with the quantum model, where ordering  does not occur even in the limit $t\to\infty$.

\bibliography{biblio}
\end{document}